\documentclass[referee]{raa}            

\usepackage{CJKutf8}

\usepackage{graphicx,times}             
\usepackage{natbib}
\usepackage{amssymb,amsmath}
\usepackage{graphicx}

\usepackage{subcaption} 
\bibpunct{(}{)}{;}{a}{}{,}

\usepackage[pagebackref=true]{hyperref}

\begin{document}

\begin{CJK}{UTF8}{gbsn}
  \title{Probing Coronal Activity Using Radio Signals Based on the 2021 superior conjunction of Mars: the Downlink Data from Tianwen-1}

   \volnopage{Vol.0 (20xx) No.0, 000--000}      
   \setcounter{page}{1}          

   \author{Yu-Chen Liu 
      \inst{1,2}
     \and De-Qing Kong
       \inst{* 1,2}
       \and Song Tan 
       \inst{3}
        \and  Zi-Han Zhao 
          \inst{4} 
           \and Zan  Wang
          \inst{1,2} 
         \and Dong-Hao Liu
          \inst{1,2}
           \and Xin-Ying Zhu
          \inst{1,2}   
            \and Yan Su
          \inst{1,2}   
             \and Hong-Bo Zhang
          \inst{1,2}  
  }

   \institute{
    National Astronomical Observatories, Chinese Academy of Sciences, Beijing 100101, China; {\it kdq@bao.ac.cn}\\
    \and
    University of Chinese Academy of Sciences, Beijing 100049, China;\\
    \and
   Leibniz-Institut f\"ur Astrophysik Potsdam (AIP), An der Sternwarte 16, Potsdam 14482, Germany;\\
    \and
    Southeast University, Nanjing, Jiangsu 210096, China
}

\vs\no
   {\small Received 2025 March 09; accepted 2026 April 05}

\abstract{ During the first superior conjunction of the Tianwen-1 Mars probe in October 2021, its downlink signal received by the Wuqing 70-m radio telescope passed within 4.53 solar radii of the Sun.
 The signal was significantly perturbed by the solar wind, providing a mechanism to probe coronal activity.
 We analyze the Doppler frequency scintillation spectrum of the solar wind within 10 solar radii to derive a characteristic frequency scintillation parameter.
 Statistical analysis indicates this parameter increases as the signal path approaches the Sun, with notable anomalies observed on October 5, 13, and 15.
 Comparisons with SOHO and SDO data reveal strong spatio-temporal correlations between these scintillation anomalies and coronal activity.
 We demonstrate that this parameter effectively identifies solar phenomena, including coronal streamers, high-speed solar wind, and coronal mass ejections (CMEs). Quantitative analysis confirms a distinct temporal correlation and delay between frequency scintillation and solar wind speed changes, validating the feasibility of spatially localizing solar activity.
}

   \authorrunning{Y.-C. Liu, D.-Q. Kong \& S. Tan, et al. }            
   \titlerunning{Solar Wind Parameter Retrieval}  

   \maketitle

\section{Introduction}\label{sect:intro}
\;\;\;\;\ \ 
When electromagnetic signals are transmitted between the Earth and other planets in the solar system,
they will pass through a series of disturbed plasma environments in deep space, 
including the ionosphere of the Earth and the target planet, 
as well as the solar wind plasma.
At this time, 
with the disturbance of the plasma environment,
the transmitted signal will also be affected by scattering, refraction and delay, 
leading to scintillation in both amplitude and frequency of
the received signal. 
When the Earth, the Sun, and the planets 
that need to communicate are
 aligned in a similar line (the superior conjunction), 
the signal will pass through the solar corona 
to the maximum extent possible. 
For at least 170 days of the 780-day Earth-Mars cycle, 
the  Sun-Earth-Probe (SEP) angle is less than 20 degrees\citep{xu+2019+effects}.
Considering the influence of solar wind,
when the interplanetary probe is at a small SEP angle,
the communication signal may be poor or even completely 
interrupted\citep{feria1997solar}.

The communication stability between deep-space probes and ground stations 
depends not only on the relative spatial positions of the Sun, 
Earth, and deep space probes 
but also on the characteristics of the channel environment 
through which radio waves propagate. 
As the SEP angle decreases, 
the distance between the signal propagation path and the Sun
gradually shortens, 
leading to an increase in the electron number density 
along the path, which  significantly intensifies 
the scattering effect of electromagnetic wave signals. 
In addition, affected by solar activities such as
coronal jets, solar flares, and coronal mass ejections (CMEs), 
the channel environment of the downlink
exhibits significant temporal dynamic disturbance 
characteristics.

This paper mainly focuses on the interaction between 
electromagnetic waves transmitted back by 
the Tianwen-1 space probe and the solar corona during the
two-week period before and after 
Mars superior conjunction 
 (i.e., a specific astronomical phase 
 where the Sun, Mars, and Earth are 
 in an approximately collinear arrangement, 
 with Mars and Earth located on either side of the Sun).
During this period, the electromagnetic wave signals 
transmitted back from the Mars probe to Earth need 
to traverse a complex propagation link composed of the 
Martian ionosphere, solar corona, and terrestrial ionosphere.
When the minimum distance from the signal propagation path 
to the Sun is $\ll 10$ solar radii($R_{\odot}$),
the typical order of magnitude of the electron number density 
in the traversed solar corona region
is $10^{9}-10^{12} cm^{-3}$\citep{MidCorona2023}, 
which is significantly higher than 
that of the Martian ionosphere 
(with a typical electron density 
range of $10^{2} - 10^{4} cm^{-3} $,
characterized by poor spatial distribution uniformity 
under the synergistic regulation of Mars' thin 
atmosphere and solar radiation\citep{Prathmesh_2025})
and the terrestrial ionosphere (excluding extreme cases 
such as geomagnetic storms, 
the peak electron density is
approximately $10^{5} - 10^{6} cm^{-3}$\citep{Jakowski_ionosph2025}).

Meanwhile, the monitoring technology 
for the terrestrial ionosphere 
has developed into a mature and systematic system.
Based on long-term accumulated observational data,
a comprehensive transmission effect correction model
has been established in relevant fields\citep{Bilitza_ionosph_2022},
which can effectively 
compensate for the disturbance impact of the terrestrial ionosphere on electromagnetic wave signals. 
Given the weak influence of the Martian ionosphere and the accurate correction of the terrestrial ionosphere's impact, 
the coronal region becomes the core influencing factor for signal transmission during this phase. 
Therefore, relevant research on solar coronal activities 
can be conducted by collecting the downlink data 
of deep space probes within the aforementioned time window.

China's Tianwen-1 Mars probe entered Mars orbit in May 2021, 
and was in the superior conjunction of Mars 
from September to October 2021. 
During this period of about one month, 
the SEP angle is less than 5 degrees, 
and the return signal link is only 4.5 solar radii 
closest to the sun received by Wuqing 70m antenna (WRT70)\citep{Kong_2022_WRT70}. 
The data during this period can be used to study the influence of solar activity on deep space communication, 
and to retrieve the physical characteristics of the perihelion corona and solar wind.
Moreover, since entering 
the 25th solar activity cycle in December 2019, 
solar activity has become more and more frequent, 
and strong solar flare events and CME events 
have erupted many times. 
Strong solar activity 
provides experimental data of 
interplanetary scintillation(IPS) in extreme cases, 
which is of great significance 
for the construction of strong solar scintillation models 
in the future\citep{xu+2019+effects,Cai2025}.

Because of the extremely high temperature 
and very thin solar corona, 
it is difficult to observe the perihelion in situ.
The remote sensing observation of the solar corona 
and solar wind by using the returned signals 
of deep space probes is an important way to obtain 
the characteristics of the solar wind near the corona 
at high latitudes. 
Scientists can also analyze the origin and dynamic propagation 
structure of CME through 
radio signals\citep{Zur2006InSituSW,Chen+2022+CME}.

Since the 1970s, scientists have used interplanetary  spacecraft's
return signals to detect solar wind characteristics
\citep{Goldstein+1969+SuperiorCO,1972ApJCronyn,1976ApJWoo,Berman+1976+DopplerNC}.
 With the mission of Tianwen-1 probe,
it is of great significance to use Tianwen-1 probe 
to observe the solar wind activities during  Solar Cycle 25. 
It is the first time that China 
has used a deep space probe outside the Earth-Moon space 
to observe and study the interplanetary medium.

Currently, there are some results on the detection of 
solar coronal activities using radio science data 
from the Tianwen-1 mission. 
\cite{MaML_2022_VLBI_SW} utilized Very Long Baseline Interferometry (VLBI) 
radio telescopes to acquire data from Tianwen‑1 
and Mars Express in October 2021. 
By identifying frequency spikes in the signals 
and performing cross‑correlation analysis, 
they detected variations in solar wind speed 
during a CME event at solar offset distances as close as approximately 2.6 solar radii.
By means of VLBI measurements, 
\cite{Wang_2023_TIW-1_DOR} utilized multiple stations  
to simultaneously receive the spacecraft’s 
Delta-Differential One-way Ranging (DOR) signals. 
They computed single-station differential phase delays (DPD),
relative variations in the total electron content (TEC), 
as well as relative fluctuations in frequency and phase,
revealed that CMEs 
can induce a sudden and significant increase 
in signal frequency in the time domain.
Utilizing VLBI observations
of data from Mars Express and Tianwen-1, 
\cite{Edwards_2025_VLBI_2021_Mars} investigated scintillation measurements of received
signals within a heliocentric distance of 40 solar radii,
which revealed that anomalies in the power-law index 
of the phase residual power spectrum are associated 
with CMEs in the near-Sun region. 
Furthermore, solar CMEs were found to induce 
spectral broadening in the received signals
\citep{Edwards_2025_VLBI_2021_Mars}.

Some researchers have also directed their focus 
to radio measurements of the Martian atmosphere, 
employing methods analogous to solar occultation of Mars 
and utilizing the geometric configuration in which 
the Martian atmosphere occults the probe 
during specific periods.
\cite{Chen_L_2023_TIW} processed and analyzed
the downlink signals from Tianwen-1 
using deep-space open-loop measurement software,
detecting radio occultation signatures such as
 amplitude diffraction fringes and 
 frequency-fitting residual uptails.
\cite{Hu_2022_TIW} employed a
one-way single-frequency radio-occultation technique,
combined with the physical properties of
the Martian ionosphere and atmosphere, 
to measure the vertical profiles of 
ionospheric and atmospheric parameters on Mars.



\section{Principles of solar wind measurements}
\label{Sec:Principle}
\begin{figure}[h]
   \centering
   \includegraphics[scale=0.75, angle=0]{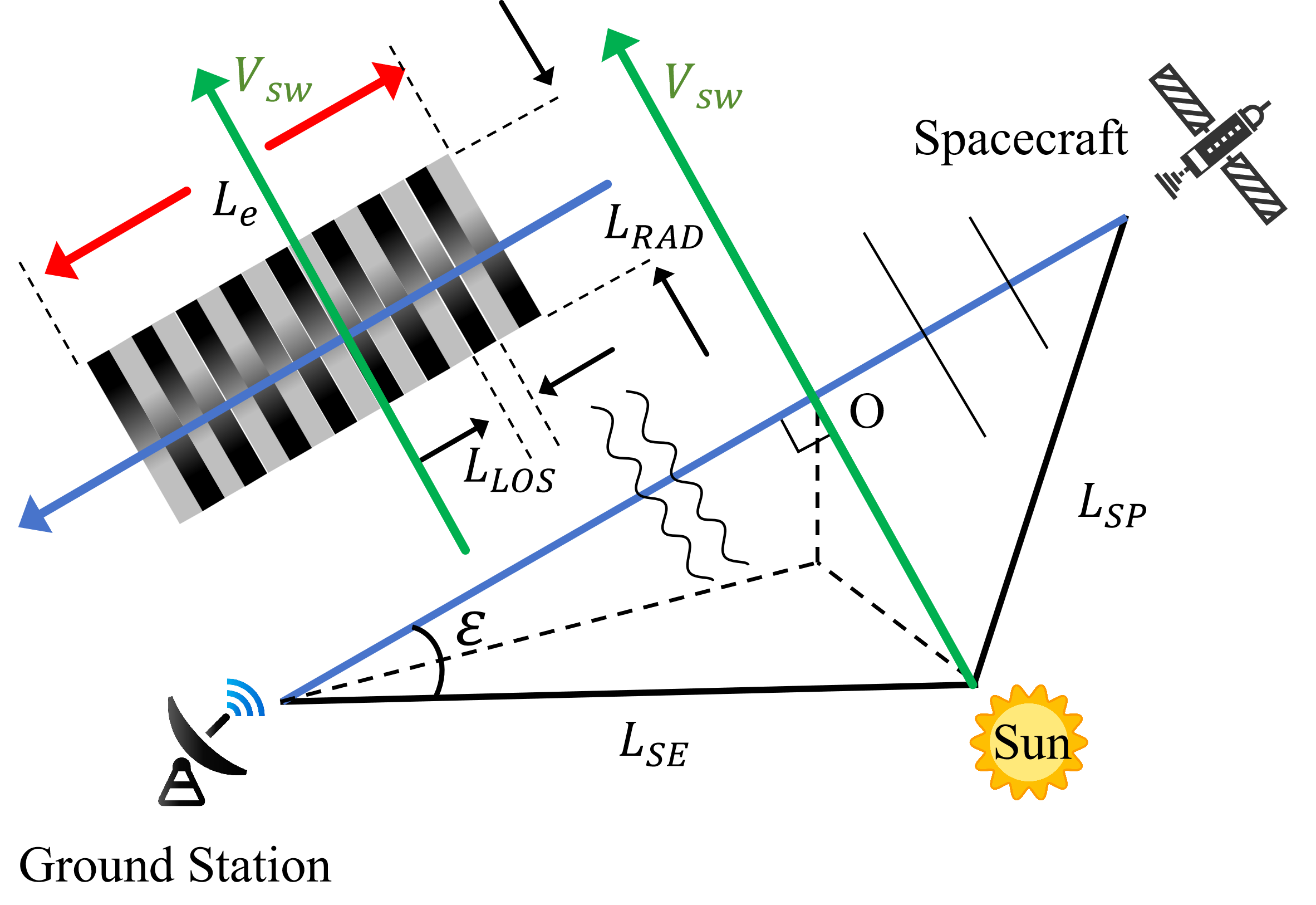}
   \caption{\quad Schematic diagram of solar wind  observation experiment based on interplanetary scintillation}
   \label{fig:IPS_theory}
   \end{figure}

\;\;\;\;\ \ 
Figure ~\ref{fig:IPS_theory} shows an experimental diagram of solar wind observations used to describe interplanetary
scintillation.
The detector transmits radio signals through the
 line of sight (LOS) and receives them at the ground station
  after passing through the $O$ point nearest to the sun. 
We denote $L_{1}=L_{SE}\cdot \cos(\angle SEP)$, 
$L_{2}=L_{SP}\cdot \cos(\angle SPE)$. 
The total distance traversed by the  downlink signal 
in interplanetary space is $L=L_{1}+L_{2}$.

As the electron density decreases with the radius of the sun, 
the degree of scattering also decreases, 
so it can be considered that the influence of 
the solar wind on the signal is mainly contributed at the point $O$. 
The distance from this point to the sun $R_{SO}$ 
is called the solar offset(SO). 
We denote $r_{SO}=R_{SO}/R_{\odot}$, $R_{\odot}=6.96\times10^{8}m$.

Doppler measurements are a common technique for radio measurements in deep space exploration. It is one of the key technologies of deep space exploration engineering to determine the precise orbit of the deep space probe by measuring the distance and its rate of change (radial velocity) in the direction of the line of sight of the ground receiving station through radio tracking measurement. 

When the electromagnetic wave signal passes through the active region of the sun, it will produce rapid phase changes, which will lead to changes in the frequency of the signal. This effect is also known as solar scintillation. This flicker change in frequency can be seen as a form of Doppler noise. Therefore, the information of frequency flicker can be obtained by estimating the Doppler noise.

\subsection{Doppler Scintillation Model}\label{FF_model}

\;\;\;\;\ \ 

Radio Doppler scintillation, 
also called frequency scintillation, 
has been widely used in solar wind retrieval. 
For example, scientists have used Doppler scintillation 
to study plasma turbulence\citep{Bruno+TurbulenceLaboratory+2013}, magnetic field perturbations\citep{Bruno+TurbulenceLaboratory+2013}, and particle acceleration\citep{Shi+2022+Acceleration} in the solar wind. In addition, scientists have used radio Doppler scintillation to study the mechanism of interaction between the solar wind and the Earth's magnetic field, as well as the impact of the solar wind on the Earth's atmosphere\citep{Liu2021SolarEffects}.

Electron density fluctuations in the solar wind are ubiquitous, and uneven density fluctuations can lead to changes in electromagnetic signals during the passage. The return signal of the detector passing through the coronal region of the sun is mainly affected by the distance parameter between the LOS and the nearest $O$ point of the sun. The observation frequency of the downlink signal emitted by the deep space probe after passing through the solar active region and reaching the ground receiving antenna can be expressed as \citep{Ptzold+2016,Wexler+2019+spacecraft}

\begin{equation}\label{f_obs_eq1}
f_{\text {obs }}(t)=f_T-\frac{V_{r e l}(t)}{c} f_T+\frac{1}{2 \pi} r_e \lambda \frac{d}{d t} \int_{\text {Earth }}^{\text {spacecraft }} n_e(s, t) d s
\end{equation}
where $f_T$ is the transmit frequency of the probe, $\lambda=c/f_{T}$ is the wavelength of the transmit carrier, and $n_e$ is the integral of path $s$ and propagation time $t$ along the LOS path. $r_e$ is the classical electron radius which has a value of $2.82\times 10^{-15}m$.
The third term in the equation ~\ref{f_obs_eq1} is the frequency fluctuation term.

In order to retrieve the solar wind parameters, the stacked slab model proposed by \cite{Wexler+2019+spacecraft} is used to model the Doppler scintillation term.

Fig~\ref{fig:IPS_theory} shows a schematic diagram of the stacked slab model. It is considered that the signal transmitted by the probe passes through a series of uncorrelated plasma density fluctuation structures along the LOS. It is believed that in each plasma slab with a thickness of $L_{LOS}$, the electron density varies with time and in the solar radial direction. It is assumed that at the same time approximately every plasma slab of thickness $L_{LOS}$ remains constant in the direction perpendicular to the solar radial direction. 
$L_{LOS}$ is called the electron concentration fluctuation correlation scale and is generally considered to be equivalent to the magnetic field autocorrelation scale representing the spacing of the magnetic flux tubes in the photosphere.

Through derivation,taking the time length of the Doppler scintillation observation experiment as $T$,
 the relationship 
between the electron density fluctuation power spectrum 
and the frequency fluctuation (FF) power spectrum can be obtained,
 with the frequency term $\nu$ expressed in units of Hz.
 \citep{Efimov2008,Wexler+2020+CoronalED}
\begin{align}
 \left|FF(\omega)\right|^{2} &\triangleq\frac{1}{T}\mathcal{F}\{\delta {f} (t)\}\mathcal{F}^{*}\{\delta {f} (t)\}\nonumber \\ 
 &=r^{2}_{e}\lambda\nu^{2}L^{2}_{LOS} \left|\delta n_{e}(\omega)\right|^{2}
\end{align}

where $\nu=\omega/2\pi$, and $\mathcal{F}\{\cdot\}$ is the Fourier transformation.
$\left|\delta n_{e}(\omega)\right|^{2}$corresponds to the power spectral density of electron concentration fluctuations

So far, the result of $\left|FF(\nu)\right|$is not only related to the solar wind parameters, but also to the wavelength $\lambda$ of the signal returned by the detector. The normalized fluctuation measure (FM) independent of the signal wavelength is defined as

\begin{equation}\label{FF_eq5}
 \left|FM(\nu)\right|^{2}= \frac{\left|FF(\nu)\right|^{2}}{\lambda^2}=r^{2}_{e}\nu^{2}L^{2}_{LOS} \left|\delta n_{e}(\omega)\right|^{2}
\end{equation}

Usually, the electron concentration is the largest at the closest position to the sun on the line-of-sight path LOS of the electromagnetic signal return. Previous studies have shown that the equivalent LOS integration path length $L_e$ can be considered as $R_{SO}/2$ before and after the $O$ point, and the number of plasma layers can be obtained as $R_{SO}/L_{LOS}$. The  radio wavelength-normalized frequency fluctuation  spectrum is then

\begin{equation}\label{FF_eq6}
 \left|FM(\nu)\right|^{2}= \frac{\left|FF(\nu)\right|^{2}}{\lambda^2}=r^{2}_{e}\nu^{2}L_{LOS}R_{SO} \left|\delta n_{e}(\omega)\right|^{2}
\end{equation}

The radio wavelength-normalized frequency fluctuation 
variance $\sigma_{FM}^2$ and 
the electron density fluctuation variance 
$\sigma_{n_{e}}^{2}$ can 
be defined\citep{Wexler+2020+CoronalED,RJain2023}, 
respectively, as integrals over the interval
$[\nu_{lo},\nu_{up}]$ as follows
\begin{equation}\label{FF_FM2}
   \sigma_{FM}^2=\int_{\nu_{lo}}^{\nu_{up}} \left|FM(\nu)\right|^{2} d\nu
\end{equation}
\begin{equation}\label{FF_Ne2}
    \sigma_{n_{e}}^{2}=\int_{\nu_{lo}}^{\nu_{up}} \left|\delta n_{e}(\omega)\right|^{2} d\nu=\frac{1}{r^{2}_{e}L_{LOS}R_{SO}}
\int_{\nu_{lo}}^{\nu_{up}}\frac{1}{\nu^2}\left|FM(\nu)\right|^{2} d\nu
\end{equation}

The upper and lower limits of the integration
are determined by the precision estimation interval 
of the power spectral density. 
A precise estimation of the power spectral density 
over a broader frequency range 
enables the extraction of more information 
regarding the structure of the solar corona.
Furthermore, the integration range must 
cover the influence interval of frequency scintillation effects
caused by coronal plasma in order to effectively 
extract information on coronal and solar wind activity 
from the frequency fluctuation power spectrum density\citep{Molera_Calves_2014_Phase_scint}.

The lower integration limit $\nu_{lo}$ is related to
the low-frequency error in the power spectral density, 
which is jointly determined by two factors: 
the cumulative duration of the observational data 
and the accuracy of slow-varying trend removal in the time domain\citep{Edwards_2025_VLBI_2021_Mars}.

The accuracy of the high‑frequency segment of the power spectral density 
is jointly constrained by the minimum time resolution 
of the observational data 
and the system noise of the spacecraft or the radio antenna. 
According to previous studies, for X‑band observations, 
the measured frequency power spectral density 
above 1 Hz is predominantly influenced by system noise.\citep{Edwards_2025_VLBI_2021_Mars} 
Under conditions of small SO, solar radiation 
causes the system‑noise band to shift 
toward lower frequencies due to 
leakage through the antenna sidelobes. 
The value of the upper integration limit $\nu_{up}$
can be determined by examining the noise‑floor diagram 
of the power spectral density.

\section{The process of Doppler observation experiment}
\label{Sec:DopplerExp}
\subsection{Introduction of Tianwen-1 Exploration Mission and Communication System}
\;\;\;\;\ \  
The working orbit of Tianwen-1 Mars probe can be divided into Earth-Mars Transfer stage and Mars capture stage. Mars parking stage, Relay Communication orbit and Remote Sensing orbit.
\begin{table}[!htbp]
    \caption{ \quad Working schedule of Tianwen-1 \citep{LiuQH2022TIW1} \citep{TIW+2021+EDL}}
  \label{tab:work_time_tianwen}
  \normalsize
    \centering
    \footnotesize
    \setlength{\tabcolsep}{36pt}
    \renewcommand{\arraystretch}{1.25}
    \begin{tabular}{lc}
        \hline
        \hline
         \textbf{Time period (CST)} & \textbf{Working Phase} \\
      \hline
            2020.07.23 -- 2020.10.09 & Launch stage \\ 
          2020.10.09 -- 2021.02.10 & Earth-Mars Transfer stage\\ 
          2021.02.10 -- 2021.02.24 & Mars capture stage \\ 
      2021.02.24 -- 2021.05.17 & Mars parking stage\\  
      2021.05.17 -- 2021.07.26& Relay Communication orbit\\ 
        2021.07.26 -- 2023.01.10& Remote Sensing orbit \\
        \hline
    \end{tabular}
    \vspace*{3ex}
    \begin{minipage}{\textwidth}
    \end{minipage}
\end{table}

From September 23 to October 23, 2021, 
427 to 457 days after the launch of Tianwen-1, 
the probe was in the Remote Sensing orbit. 
During these 30 days, the SEP angle formed 
by the Sun-Earth-probe was less than 5 degrees, 
the closest distance (SO) of the path of 
the radio signal returned by Tianwen 1 
from the heliocentric point was less than 
19 solar radii $R_{\odot}$, 
the distance between the probe 
and the earth was about 400 million kilometers, 
and the one-way travel time was about 22 minutes\citep{Geng+2018}.

The uplink and downlink TT$\&$C (Tracking, 
Telemetry and Command) communication of the detector 
will be affected during the superior conjunction.
Table ~\ref{tab:work_time_tianwen}
shows the working phase of the Tianwen-1 deep space probe.
During the superior conjunction in 2021, 
Tianwen-1 will be in the remote sensing orbit.
For the safety of the device,
Tianwen 1 does not transmit scientific data to the ground, 
but only uses a low-gain antenna 
to transmit TT$\&$C information to the ground
with the downlink EIRP $\geq 21$dBW 
(Low Gain Antenna within the range of $\pm50^{\circ}$).

\subsection{Introduction of Ground Receiving System}
Compared with previous lunar exploration missions, the first Mars exploration mission in China has the following characteristics:
\begin{enumerate}
    \item The signal transmission distance is longer. 
    During the superior conjunction of Mars, 
    the distance between Mars and the Earth 
    reached about 400 million kilometers. 
    Such long-distance signal transmission 
    will bring greater free propagation loss. 
    The maximum loss of Mars communication path 
    is 60 dB higher than that of lunar earth communication,
    and the maximum loss reaches 81 dB\citep{LiuJJ2018};
    \item Long-distance signal transmission 
    leads to a maximum signal delay of 22 min, 
    which puts forward higher requirements 
    for the ability of space and earth measurement and control
    and the ability of autonomous operation 
    of the Mars probe\citep{Geng+2018};
    \item The ground receiving system 
    needs to process the multi-payload data 
    of  the Mars probe, 
    and involves the data synthesis of multiple antennas, 
    which greatly increases the complexity of the ground data 
    processing;
    \item The lack of prior knowledge of 
    the Earth-Mars channel environment 
    increases the difficulty of 
    signal tracking and demodulation in extreme cases.
\end{enumerate}

The ground receiving system adopts the working mode of multi-station antenna array receiving. The X-band signal transmitted by the detector is received by the antenna and then sent to the low noise amplifier connected to the antenna feed output port. The signal is subjected to low noise amplification and down-conversion to obtain the pre-detection data stored in the local antenna.\citep{LiuJJ2018}
In this paper, Wuqing 70m antenna (WRT70) is used for the experiment.\citep{Zhang+Mars+2021,Kong_2022_WRT70}

\subsubsection{Observation Experiment \& Data Preparation}
\label{sec:occlutation}

During the superior conjunction of Mars in 2021, 
the Tianwen-1 probe no longer transmits scientific data, 
but transmits X-band downlink TT$\&$C data through a low-gain antenna. 
From September 22, 2021 to October 20, 2021, 
WRT70 completed the reception of 
the downlink data of the low-gain antenna transmission signal of Tianwen-1 detector. 
In this paper, the solar wind observation experiment is carried out 
by using the data of the downlink TT$\&$C link 
whose distance is less than 10 solar radii 
(from October 1 to October 15, 2021), 
and the Doppler frequency scintillation and 
amplitude scintillation data
are obtained by using the improved Doppler estimation algorithm 
and the improved detrending algorithm.

The sampling frequency of the data is 6.25 MHz or 1.25MHz and 8-bit quantization is taken. Due to the influence of Doppler frequency shift, the main frequency of the signal is not strictly zero intermediate frequency. The observation data is recorded in the following table.

\begin{table}[h]
    \caption{\quad Observation data during the superior conjunction of Mars in 2021}
    \label{tab:IPS_experiment_plan}
    \centering
    \footnotesize
    \setlength{\tabcolsep}{4pt}
    \renewcommand{\arraystretch}{1.2}
    \begin{tabular}{lcccccc}
        \hline
        \textbf{Date}  & \textbf{Begin of observation} &\textbf{End of observation} &\textbf{SO} & \textbf{Sampling rate}&\textbf{File Size}\\
       MM.DD  & \textbf{UTC} & \textbf{UTC}& $R_{\odot}$ &  \textbf{MHz}&\textbf{GB}\\
       \hline
        \hline
                \multicolumn{6}{c}{Ingress}\\
                  \hline
          10.01& 00:14:35 & 04:05:42& 9.24-9.04  &1.25&32.3\\ 
       10.02& 00:14:59& 03:59:58 &8.05-7.86  &1.25&31.4\\ 
       10.03& 00:55:59&07:40:00  &  6.84-6.55 &1.25&56.0\\  
        10.04& 00:24:59 &03:19:58& 5.75-5.61 & 1.25&24.4\\ 
        10.05& 00:04:59 & 03:35:00&4.66-4.53 &1.25&29.3\\   
       \hline
    \multicolumn{6}{c}{Egress}\\
         \hline
         10.12& 01:01:59 &02:48:53 &5.32-5.41 &1.25&14.9\\    
       10.12& 07:59:59 &10:00:00&  5.63-5.73 & 6.25&83.9\\     
        10.13& 01:24:59&08:30:00 &  6.44-6.76 & 6.25&297\\           
           10.14 & 01:19:59& 02:55:57& 7.59-7.68& 6.25&67.1\\    
        10.14 & 02:59:59&08:30:00 & 7.68-7.92& 1.25&46.1\\     
       10.15 & 00:54:59&08:29:59 & 8.76-9.14& 1.25&63.6\\ 
        \hline
    \end{tabular}
    \vspace*{3ex}
    \begin{minipage}{\textwidth}
    \end{minipage}
\end{table}

The total amount of observed data is 746 GB, 
of which the minimum SO distance on October 5 is 4.66 solar radii, 
corresponding to a SEP angle of 1.42 degrees. On October 8, 2021, 
the SEP angle reached a minimum of 0.65 degrees. 
From Earth's perspective, Mars appeared to the left of 
the Sun before 2021 October 8, and to the right of 
the Sun after 2021 October 8. 
In this paper, we generally use "ingress" or "left" to 
refer to the period before Mars enters superior conjunction,
and "egress" or "right" to refer to the period 
after Mars exits superior conjunction.

\subsection{Scintillation Measurement \& Data Processing}
\;\;\;\;\ \
Near the solar active region, the density of solar wind plasma increases sharply with the decrease of radial distance, and the plasma flow has strong turbulence and uneven distribution, and there is a possibility of instantaneous disturbance (i.e. coronal mass ejection, CME). When the detector signal passes through the solar active region, it will produce spectrum spreading, frequency scintillation, group delay and other phenomena, which will lead to the loss of lock in the demodulation of the phase-locked loop in the ground receiver.

Because the downlink received signal is often severely disturbed during the superior conjunction of Mars, it is generally not possible to directly use the PLL for frequency estimation. Some scholars divide the data per second into data blocks, and find the extreme point of frequency by Discrete Fourier Transform(DFT) to estimate the frequency.\citep{Imamura+2005+NOZOMI, Imamura+2014+StrongScatter, Shota+2022+FastSlow}. This method can quickly get the preliminary results of the frequency change, but the resolution is low in the frequency domain. In the subsequent calculation of the power spectrum, it is difficult to obtain the information of the high frequency part.

The downlink signal returned by Tianwen-1 received by the ground observation station can be expressed as
\begin{equation}\label{f_obs_eq11}
f_{\text {obs }}(t)=f_T-\frac{V_{r e l}(t)}{c} f_T+\delta f
\end{equation}

The second term in Equation (~\ref{f_obs_eq11})
arises from the Doppler shift 
caused by the relative motion 
between the deep-space probe and the ground receiving station.
This varies relatively gradually 
compared to the frequency scintillation 
induced by solar activity, 
which exhibits fluctuations
with a resolution of less than a second.

Since the Doppler frequency shift 
constitutes a long-term effect, 
selecting an appropriate method 
for removing its trend can effectively 
reduce errors in the low-frequency portion 
of the FF power spectrum, 
thereby enabling a more accurate estimation 
of the spectrum. 
This data-processing step directly 
affects the selection of 
the low-frequency integration limits $\nu_{lo}$ in Equation (\ref{FF_FM2}) and (\ref{FF_Ne2}) .

\cite{Liu_2023} developed 
a multi‑level iterative correction algorithm 
based on Kalman filtering. 
In accordance with the characteristics of 
solar activity interference, this algorithm 
corrects phase jumps caused by solar activity 
in signals returned by the Tianwen‑1 spacecraft during Mars superior conjunction. 
This enables the PLL to operate stably even 
under extreme conditions, thereby providing 
a more accurate frequency estimation of the returned signal.

\begin{figure}[!ht]
    \centering  
    \includegraphics[width=0.9\textwidth]{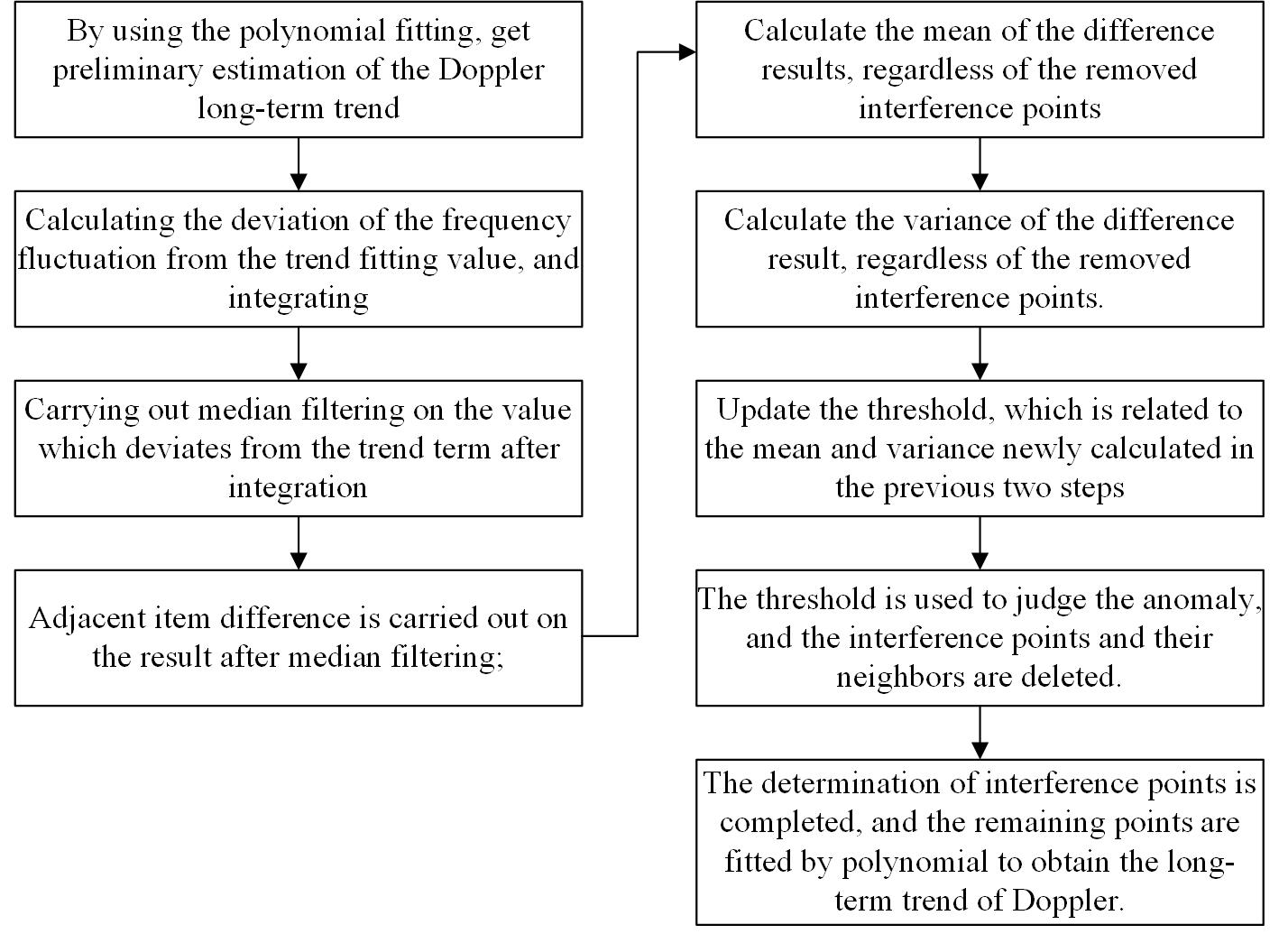}
    \caption{\quad Schematic Diagram of Doppler Change Trend Fitting Algorithm} \label{detrend_method}
\end{figure}

After the received signal passes through the PLL, 
frequency scintillation containing the Doppler shift 
can be extracted. 
During periods of severe frequency scintillation, 
\cite{Liu_2023} found that using direct polynomial fitting
to remove the relatively slow-varying Doppler trend term 
is significantly affected by rapid frequency scintillation, 
which compromises the accuracy of trend removal. 
Consequently, they proposed an error-correction-based trend removal method. 
Building upon this method, the concept of multi-level iteration was introduced for further optimization, 
enabling more refined extraction of the Doppler trend term.
The detailed algorithmic procedure is illustrated in Fig. \ref{detrend_method}.

First, a coarse extraction of the trend term 
is performed using direct polynomial fitting. 
After subtracting this coarsely extracted trend 
from the time‑series data, median filtering is applied. 
The deviation of each data point from the median‑filtered 
result is then computed. 
Based on the mean and variance of the data,
a detection threshold is set to identify the coordinates 
of anomalous points. 
Since frequency fluctuations caused by coronal plasma disturbances 
do not manifest as single‑point jumps, 
this step requires the removal of 
not only the anomalous points themselves 
but also the data points 
within a neighboring time interval around each anomaly. 
The remaining data are then subjected to 
a new polynomial fitting.

The steps described above constitute one complete iteration 
of the trend‑fitting process. 
This algorithm can be iterated multiple times 
until the difference between successive fitting results 
falls below a predefined threshold, 
thereby completing the final extraction of 
the Doppler trend term. After eliminating the Doppler frequency shift, 
the instantaneous Doppler scintillation $\delta{f}(t)$ 
can be obtained.


In the polynomial fitting described above, 
the selection of polynomial degree is also 
a critical consideration. 
The work by \cite{Molera_Calves_2021_method} 
provides a detailed discussion on the choice of 
polynomial degree for the trend terms of 
both frequency residuals and phase residuals. 
For frequency residuals, the polynomial degree 
typically falls within the range of 6 to 9, 
although in rare cases,
a polynomial of up to degree 15 may be required. Regarding the selection of polynomial degree for fitting, 
the results from \cite{Edwards_2025_VLBI_2021_Mars} are adopted.
Observations within 18.7 solar radii are fitted using a 9th‑degree polynomial. 
Over‑fitting would lead to the removal of 
low‑frequency contributions, 
while under‑fitting could result in the 
inaccurate estimation of low‑frequency components 
due to inadequate high‑precision Doppler compensation\citep{Molera_Calves_2021_method,Edwards_2025_VLBI_2021_Mars}.

The next step is to determine 
the integration interval $[\nu_{lo},\nu_{up}]$ 
for both the radio wavelength-normalized frequency fluctuation variance(~\ref{FF_FM2}) and the electron density fluctuation variance(~\ref{FF_Ne2})

\cite{Shota+2022+FastSlow} used Akatasuki's observations at SO 1.5-8.9 $R_{\odot}$ to divide the daily observations into 2000 seconds intervals, and the final spectral resolution was 1.31 seconds.
\cite{Wexler+2019+spacecraft} used 4000s data segments for observation. The time length for \cite{Ma+2021} to draw FF spectrum is about 1000 seconds. \cite{Bruno+2009+Coord} pointed out that the time resolution of FF spectra needs to be less than local ion gyro-period of the plasma, and the observation time needs to be relatively long\citep{Bruno+TurbulenceLaboratory+2013}.
In this chapter, the time resolution of the power spectrum is 0.005 s, 
taking 3272.8 s as a period of time, considering the limitation of the effective data observation time and the previous processing methods.

After the FF power spectrum is obtained, 
the formula (~\ref{FF_FM2}) shows that the corresponding  $\sigma_{FM}$
can be obtained after the integration interval 
$[\nu_{lo},\nu_{up}]$ is determined. 
The commonly used integration intervals are 1–28 mHz and 0.5–100 mHz.\citep{wexler2020phD}
 \cite{Liu_2023}. Previous researchers generally take the average of these two integration intervals (one narrow and one wide) 
to obtain the estimated value of  $\sigma_{FM}$.

\begin{figure}[!ht]
    \centering
    \includegraphics[width=0.8\textwidth]{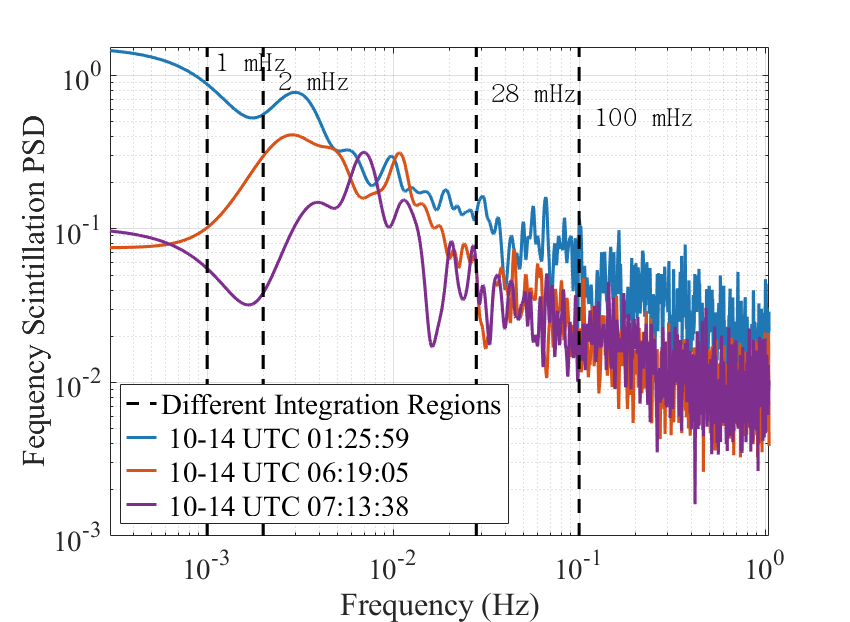}
    \caption{\quad Comparison of Multiple PSD Curves 
    and Integration Regions on 2021 Oct 14}
    \label{fig:PSD Curves and Integration}
\end{figure}
In order to select an appropriate integration interval 
for the  $\sigma_{FM}$ estimation in this observation experiment, 
several PSD curves observed on October 14, 
when solar activity is relatively quiet, 
are plotted as Figure ~\ref{fig:PSD Curves and Integration}. 
A comparison with the commonly used integration interval 
is also conducted. The theoretical upper limit of 
the integration is related to the sampling rate
 of the observed data. 
 Although the sampling rate in this experiment 
 is relatively high (200 Hz), 
 the upper limit of integration 
 is still constrained by system noise. 
 Therefore, the upper limit for the relatively 
 wide-range integration interval 
 is set at 100 mHz, 
 while for the narrow-range integration interval, 
 it is set at 28 mHz, 
 consistent with previous research findings.

Regarding the lower limit of integration, 
the cumulative observation time for 
the single-segment data is 3272.8 seconds, 
corresponding to a theoretical optimal
lower integration limit of 0.3 mHz. 
However, since multi-segment averaging 
is required for PSD estimation, 
the actual theoretical optimal low-frequency resolution 
is slightly less than 1 mHz. 
Furthermore, the accuracy of the power spectrum estimation
in the low-frequency band 
is constrained by the precision of 
the Doppler trend removal algorithm. 
Research by \cite{Liu_2023} indicates that the accuracy 
of the multi-level iterative correction algorithm 
for detrending is approximately 2 mHz, 
and the accuracy in the low-frequency region below 2 mHz
may be influenced by the choice of polynomial fitting degree.
Ultimately, in this observation, 
the lower limit for both the wide-range and narrow-range 
integration intervals is set at 2 mHz.

Returning to  Figure~\ref{fig:PSD Curves and Integration}, 
with the upper integration limit set at 100 mHz, 
the PSD curve has not yet reached the noise floor. 
Thus, the choice of the upper integration limit 
is reasonable. However, if 1 mHz were selected 
as the lower integration limit, 
it can be seen that the PSD curves 
corresponding to the observation start times of 
01:25:59 UTC and 07:13:38 UTC exhibit a pattern 
in the low-frequency band of first decreasing, 
then increasing, and then decreasing again. 
Yet, according to the research by \cite{efimovOuterScaleSolarwind2002}, 
the theoretical FF spectrum in the 
low-frequency band either flattens out before declining 
or shows a trend of first rising and then falling. 
This aligns with the trend observed after 2 mHz. 
This demonstrates that selecting 2 mHz as the 
lower integration limit is consistent with the observations.

\begin{equation}
\begin{aligned}
&  \sigma_{FF_{1}}^2=\int_{2mHz}^{28mHz} \left|FF(\nu)\right|^{2} d\nu \\ \\
&  \sigma_{FF_{2}}^2=\int_{2mHz}^{100mHz} \left|FF(\nu)\right|^{2} d\nu 
\end{aligned}
\end{equation}

The normalized RMS(root-mean-square) frequency fluctuation measure is obtained according to (~\ref{FF_eq5}).
\begin{equation}\label{eqs_fm}
    \sigma_{FM}=\frac{\sqrt{(\sigma_{FF_{1}}^2+\sigma_{FF_{2}}^2)/2}} {\lambda} 
\end{equation}

\subsection{Observation Results}

\;\;\;\;\ \
Figure ~\ref{fig:PSD Curves and Integration}
presents the arrangement of the FF scintillation spectrum 
results for a day with relatively quiet solar activity. 
On 2021 October 14, the Tianwen-1 spacecraft 
was located on the right-hand side of the Sun (as seen from Earth). 
As time progressed, the SO distance along the 
signal transmission path for the probe's downlink signal 
gradually increased, 
and the FF spectrum exhibited a decreasing trend over time. 
This variation trend is also consistent with theoretical expectations.
\begin{figure}[htbp]
    \centering
    \begin{subfigure}{0.492\textwidth}
        \centering
        \includegraphics[width=\textwidth]{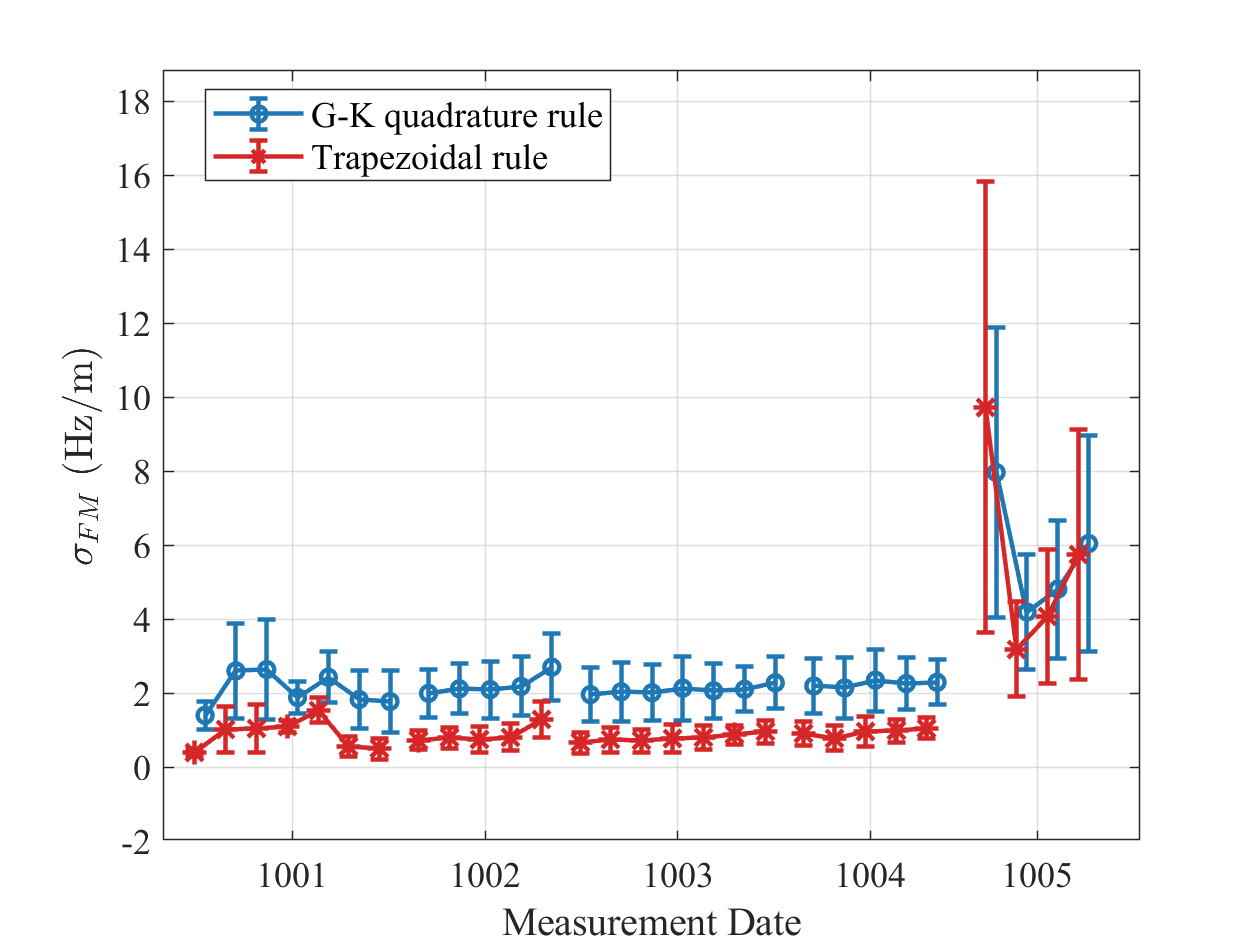}
        \caption{}
        \label{fig4:sub1}
    \end{subfigure}
    \hfill
    \begin{subfigure}{0.498\textwidth}
        \centering
        \includegraphics[width=\textwidth]{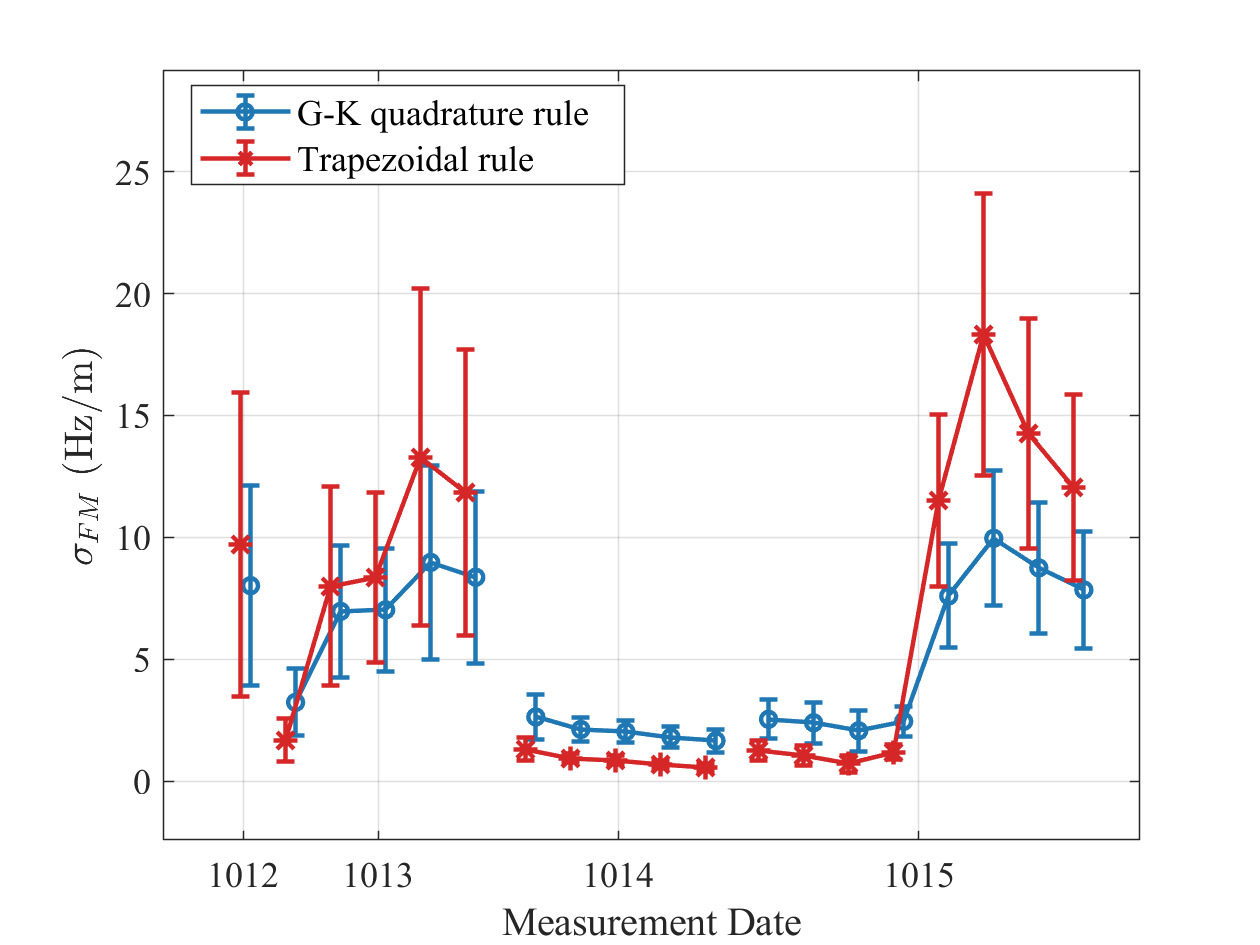}
        \caption{}
        \label{fig4:sub2}
    \end{subfigure}
    \caption{The normalized RMS frequency fluctuation measure(FM) Statistics of Frequency Fluctuations  $\sigma_{FM}$ during the 2021 Mars Solar Conjunction: Comparison between Two Integration Methods.
    (a):Comparison of   $\sigma_{FM}$ Obtained by Two Integration Methods in the ingress phase.
    (b):Comparison of   $\sigma_{FM}$ Obtained by Two Integration Methods in the egress phase.}
    \label{fig:FM Statistics}
\end{figure}

Combining with Figure ~\ref{fig:PSD Curves and Integration}, 
it can be clearly seen that 
the FF power spectrum exhibits the characteristics 
of the Kolmogorov spectrum, 
with the spectrum fluctuating rapidly 
as the frequency increases. 
To obtain the calculation result of Equation(~\ref{eqs_fm}), 
the traditional trapezoidal rule may 
introduce certain estimation errors. 
Therefore, the Gauss–Kronrod quadrature rule 
is introduced here to perform a more accurate 
numerical integration tailored to the features of 
the Kolmogorov spectrum. 
Figure ~\ref{fig:FM Statistics} compares the results of the two integration 
methods and presents the $\sigma_{FM}$ during 
the observation period.
These findings demonstrate the consistency 
between the two integration methods. 
This validates the use of the improved method
for further result analysis.

For the two different integration methods,
the $\sigma_{FM}$ values during the observation period 
were statistically analyzed, 
and Figure ~\ref{fig4:sub1} and ~\ref{fig4:sub2} was plotted. 
It should be noted that the connected data points 
in Figure ~\ref{fig:FM Statistics} represent data 
from the same day. 
To clearly illustrate the trend of FM variation  $\sigma_{FM}$ in the figure, 
the spacing between consecutive  $\sigma_{FM}$ data points 
on the x-axis is not proportional to the actual time intervals. 
Instead, the data are arranged 
at equal intervals along the x-axis 
in chronological order to generate Figure ~\ref{fig:FM Statistics}.

It can be seen that both the data during ingress and egress
show that there is a certain relationship 
between the $\sigma_{FM}$ (or PSD of FF, i.e. see figure ~\ref{fig:PSD Curves and Integration}) and the SO radial distance. 
In general, the spectra are arranged from top to bottom as the radial distance decreases. According to the theory of turbulent spectrum, it can be concluded that the fluctuation of electron density in the solar wind decreases with the radial direction.

This result is also broadly 
consistent with Equations ~\ref{FF_FM2} and ~\ref{FF_Ne2}.
Since the value of $L_{Los}$ is not known in advance, 
it is not feasible to accurately determine the 
electron density fluctuations. 
According to the expression in Equation ~\ref{FF_Ne2}, 
it is reasonable to assume that over a certain period, 
$L_{Los}$ does not vary with the SO distance. 
Under this assumption, 
the relative variation in electron density 
fluctuations $\sigma_{ne}$ is consistent with the 
relative variation in $\sigma_{FM}$.

In the ingress phase, 
as time progresses,
the SO distance decreases. 
When the daily data are averaged into a single point, 
the  $\sigma_{FM}$ values generally follow a gradually increasing trend. 
Meanwhile, the intra-day data 
on October 2, 3, and 4 also 
conform to the pattern of $\sigma_{FM}$ increasing over time.
During the egress phase, 
the data from October 12 to 15 (four days) 
generally exhibited a trend of 
decreasing   $\sigma_{FM}$
with decreasing SO. 
However, during certain time intervals 
on October 13 and 15,
$\sigma_{FM}$ showed anomalous increases.

During periods of relatively quiet solar activity 
(in the absence of high-speed solar wind), 
Researchers derived 
the relationship between $\sigma_{FM}$ and
electron density $n_e$ based on 
multi-year observational data from 
different spacecraft and 
verified its validity within the mid-corona 
range\citep{wexler2020phD,wexler2022slow}.
($1.5 R_{\odot}\leq R_{SO} \leq 6R_{\odot}$)
\begin{equation}\label{eqs:sigma_FM_eq1}
\sigma_{FM}=9.69 \times 10^{-12} n_{e}\sqrt{r} 
\end{equation}
where $n_e$ is the electron density 
at point $O$ in Figure~\ref{fig:IPS_theory}.

At closer solar distances, coronal jets or 
solar wind bursts are hardly avoidable 
even under quiet or low solar activity conditions. 
Therefore, an additional correction 
term must be introduced into the formula\cite{wexler2022slow}.

\begin{equation}\label{eqs:sigma_FM_eq2}
  \sigma_{FM}=9.69 \times 10^{-12} n_{e}\sqrt{r} +49.7r^{-3.2}
\end{equation}

\begin{figure}[!ht]
    \centering
    \includegraphics[width=0.95\textwidth]{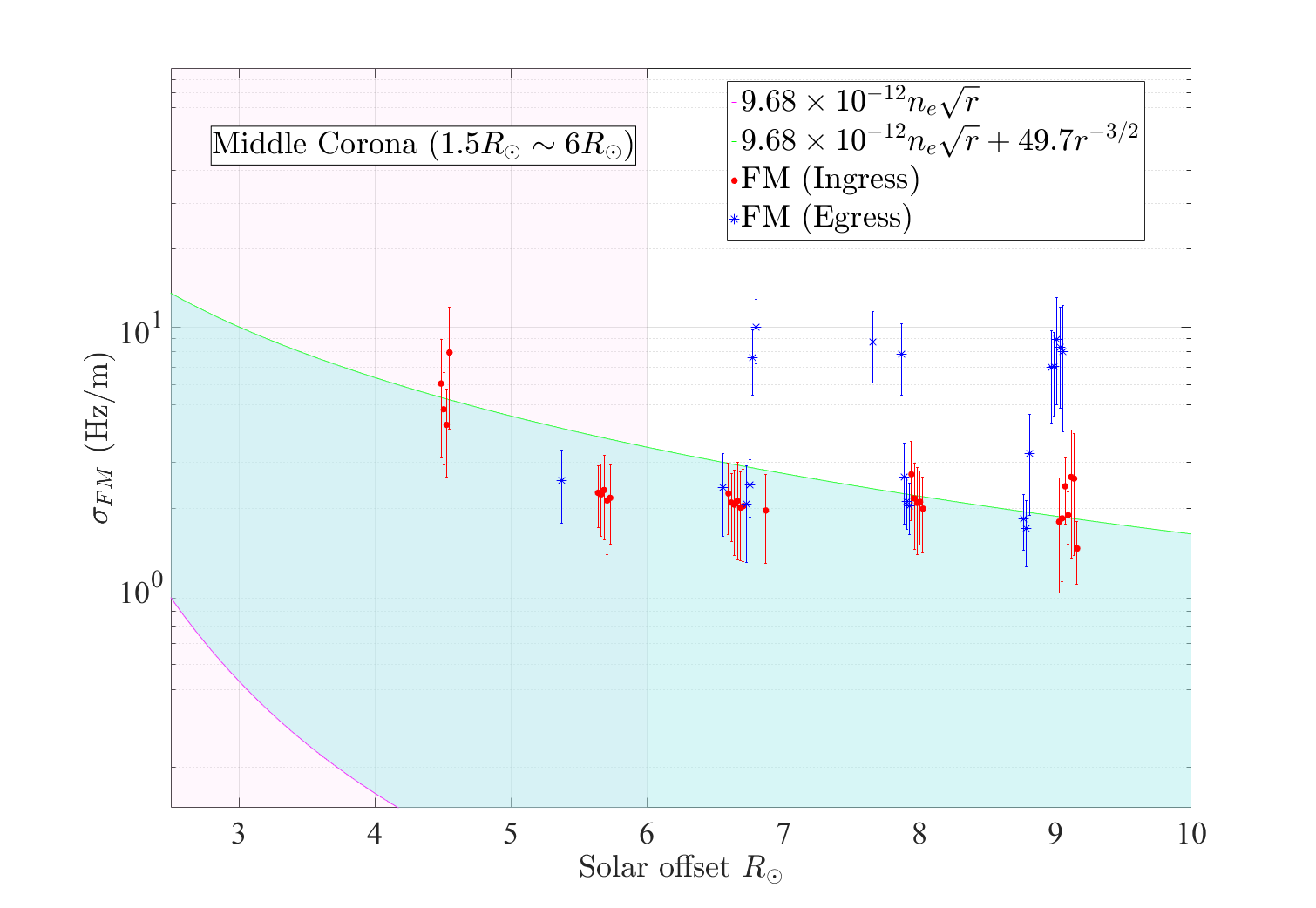}
    \caption{\quad The normalized RMS frequency fluctuation measure Variation with Solar offset during the 2021 Mars Solar Conjunction}
    \label{fig:solar_radii_FM}
\end{figure}

 The ingress data in Figure ~\ref{fig4:sub1}
 corresponds to the red data points 
 and red error bars in Figure ~\ref{fig:solar_radii_FM}., 
 while the egress data in Figure ~\ref{fig4:sub2}
 corresponds to the blue data points and blue error bars 
 in Figure 5. 
 The vertical axis of the plot is in logarithmic scale, 
 and the horizontal axis represents solar offset 
 in units of solar radii.The results are 
 also compared with the curves from 
 Equations ~\ref{eqs:sigma_FM_eq1} 
 and ~\ref{eqs:sigma_FM_eq2}.

In Figure ~\ref{fig:solar_radii_FM}, 
the blue asterisk represents the observation data 
observed in the ingress phase, 
and the red circle represents the 
observation data during egress. 
The observation of the data is in good agreement 
with the result of (~\ref{eqs:sigma_FM_eq2}).

\section{Results Analysis}
\;\;\;\;\ \
Combined analysis of the data 
from  Figures~\ref{fig:FM Statistics} and ~\ref{fig:solar_radii_FM} suggests that 
solar activity on October 5, 13, and 15 was potentially 
subject to anomalous perturbations. 
During the other days, solar activity 
along the propagation path of Tianwen-1's downlink signal 
remained relatively calm.

In the model described in Section ~\ref{Sec:Principle},
FM variations  $\sigma_{FM}$ are primarily attributed 
to contributions 
from the "$O$ point" along the signal return path (Figure ~\ref{fig:IPS_theory}). 
The $O$ point migrates slowly with the orbital motion of Mars,
and no in‑situ probe is permanently located at this position.
SOHO operates near the Sun–Earth L1 Lagrange point, 
and during Mars superior conjunction, 
the viewing geometry of Mars from SOHO closely aligns with that from Earth. 
Hence, SOHO observations provide valuable references 
for characterizing the physical properties at the $O$ point. 
Furthermore, during the period when Mars is occulted 
by the Sun, since the Solar Dynamics Observatory (SDO) 
operates in a geosynchronous orbit, 
point $O$ also lies precisely within 
its field of view and is essentially aligned 
with the viewing angle observed by SOHO.

To investigate the causes of the data anomalies 
observed on the aforementioned days, 
we compared observations from SOHO and SDO 
to assess the intensity of solar activity 
along Tianwen‑1's line‑of‑sight (LOS) direction.
Through comparative analysis, 
it was found that these data anomalies 
are related to coronal streamers, 
high-speed solar wind streams, and CMEs. 
An elaboration on each of these phenomena 
is presented in the following sections.
\subsection{CMEs}
\;\;\;\;
\begin{figure}[htbp]
    \centering
    \begin{subfigure}{0.48\textwidth}
        \centering
        \includegraphics[width=\textwidth]{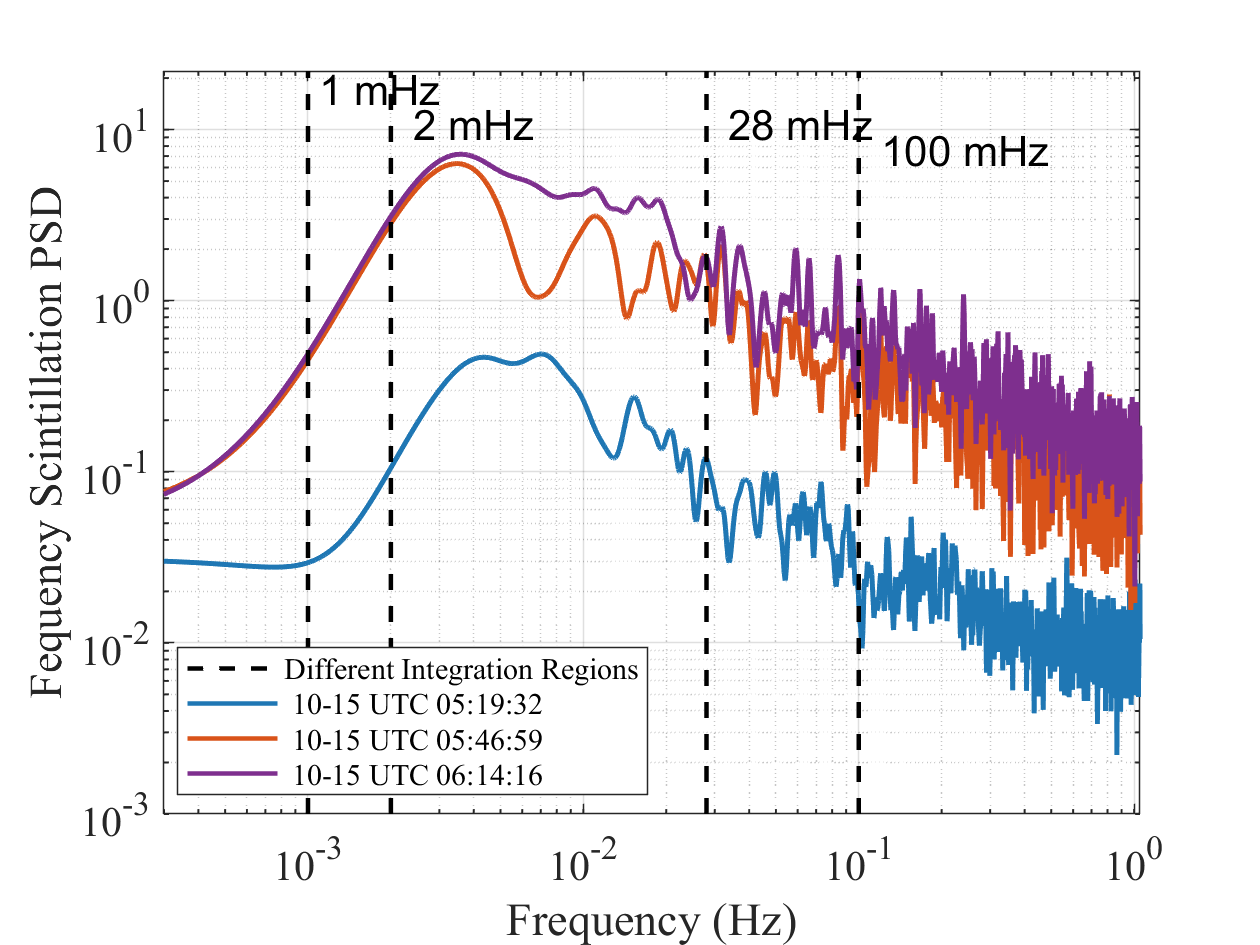}
        \caption{}
        \label{subfig:1015_PSD}
    \end{subfigure}
    \hfill
    \begin{subfigure}{0.48\textwidth}
        \centering
        \includegraphics[width=\textwidth]{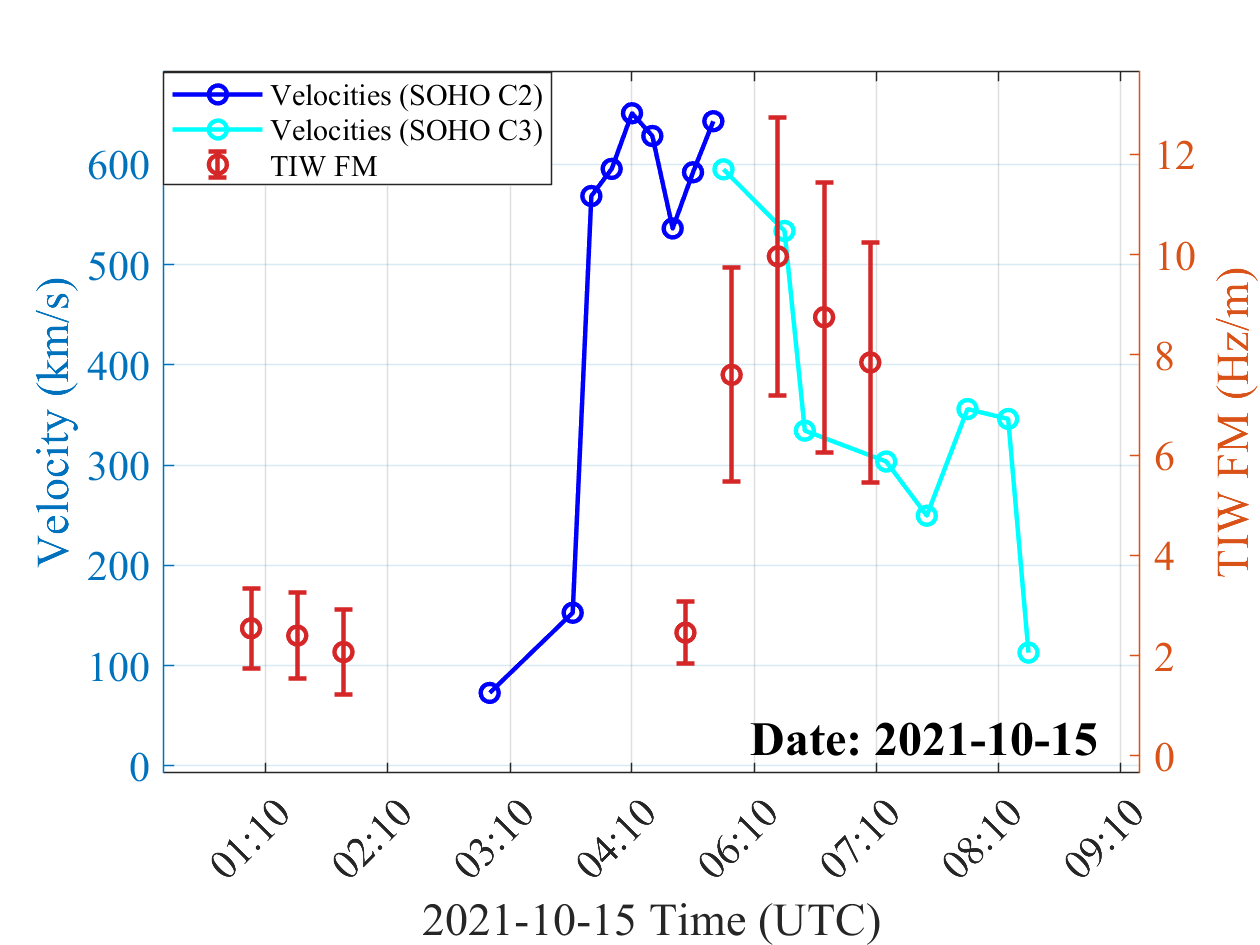}
        \caption{}
        \label{subfig:1015FM}
    \end{subfigure}
     \hfill
    \begin{subfigure}{0.8\textwidth}
        \centering
        \includegraphics[width=\textwidth]{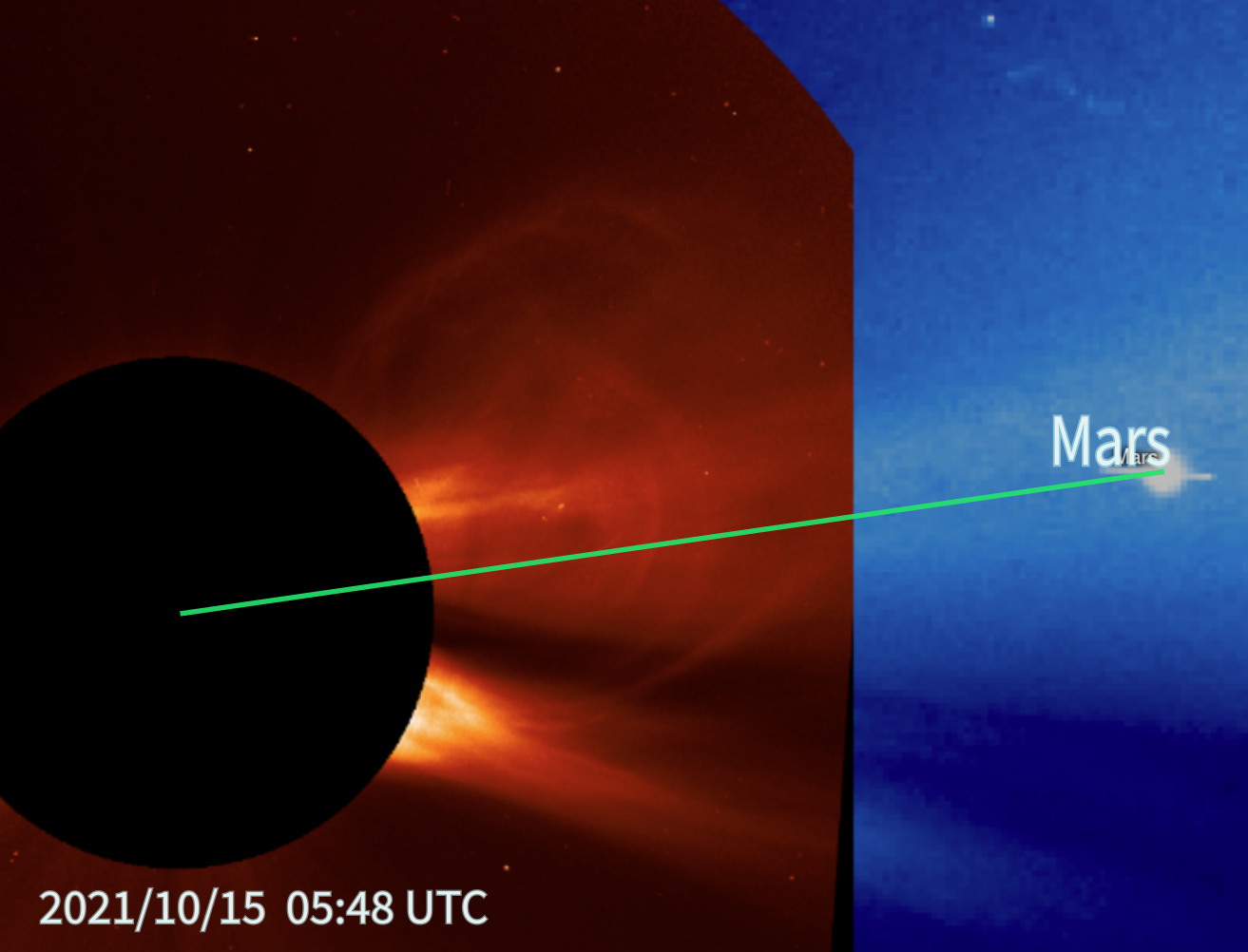}
        \caption{}
        \label{subfig:1015_SOHO}
    \end{subfigure}
    \caption{ Relationship between FM fluctuations   $\sigma_{FM}$and CMEs.(a):The frequency scintillation power spectrum observed on 2021 October 13.
    (b):Comparison between solar wind speed variations and variations in the  $\sigma_{FM}$. (c):A CME eruption observed on the right-hand side of the Sun on October 15,
     2021.}
    \label{fig:1015}
\end{figure}
It can be seen from the frequency scintillation spectrum 
and   $\sigma_{FM}$ (Figure~\ref{subfig:1015_PSD} and ~\ref{subfig:1015FM}) on 2021 October 13 that the data on that day 
were anomalous.

Since October 13 was already in the egress phase, 
the frequency scintillation spectrum should
have shown a progressively decreasing pattern over time
—that is,   $\sigma_{FM}$ should have been decreasing. 
This suggests that there may have been 
anomalous solar activity on that day.

Signatures of CME eruptions are captured 
in optical images obtained by 
the SOHO LASCO C2 or C3 coronagraphs. 
Using the central difference method, 
the solar wind velocity at the eruption location 
can be estimated based on the CME's measured 
displacement and the corresponding time 
stamps from SOHO.

In the lower panel of Figure ~\ref{fig:1015}, 
an image captured by the SOHO C2\&C3 coronagraph on 
October 15 is presented. 
The green line indicates the approximate direction 
toward Mars, corresponding to a SO distance 
of approximately 8.95 solar radii at that moment.
In the  upper right panel, 
the solar wind velocity obtained from 
SOHO C2 and C3 images via 
the central difference method is shown, 
together with a comparison to the corresponding 
$\sigma_{FM}$ variations.

A maximum in $\sigma_{FM}$ was observed at approximately
06:14 UTC, while SOHO measurements showed that the 
CME reached its peak velocity during the interval 
from around 04:48 to 05:36 UTC. As shown in Figure ~\ref{subfig:1015FM},
the variations in the solar wind velocity estimated 
from SOHO optical images exhibit a good overall 
trend agreement with the  $\sigma_{FM}$ anomalies, 
although a noticeable time lag 
is present between the two measurements. 
A quantitative analysis of this time delay,
together with that observed in other events, is provided in 
Section ~\ref{sec:corr_velo}.

\subsection{High-Speed Solar Wind Streams}
\begin{figure}[htbp]
    \centering
    \begin{subfigure}{0.51\textwidth}
        \centering
        \includegraphics[width=\textwidth]{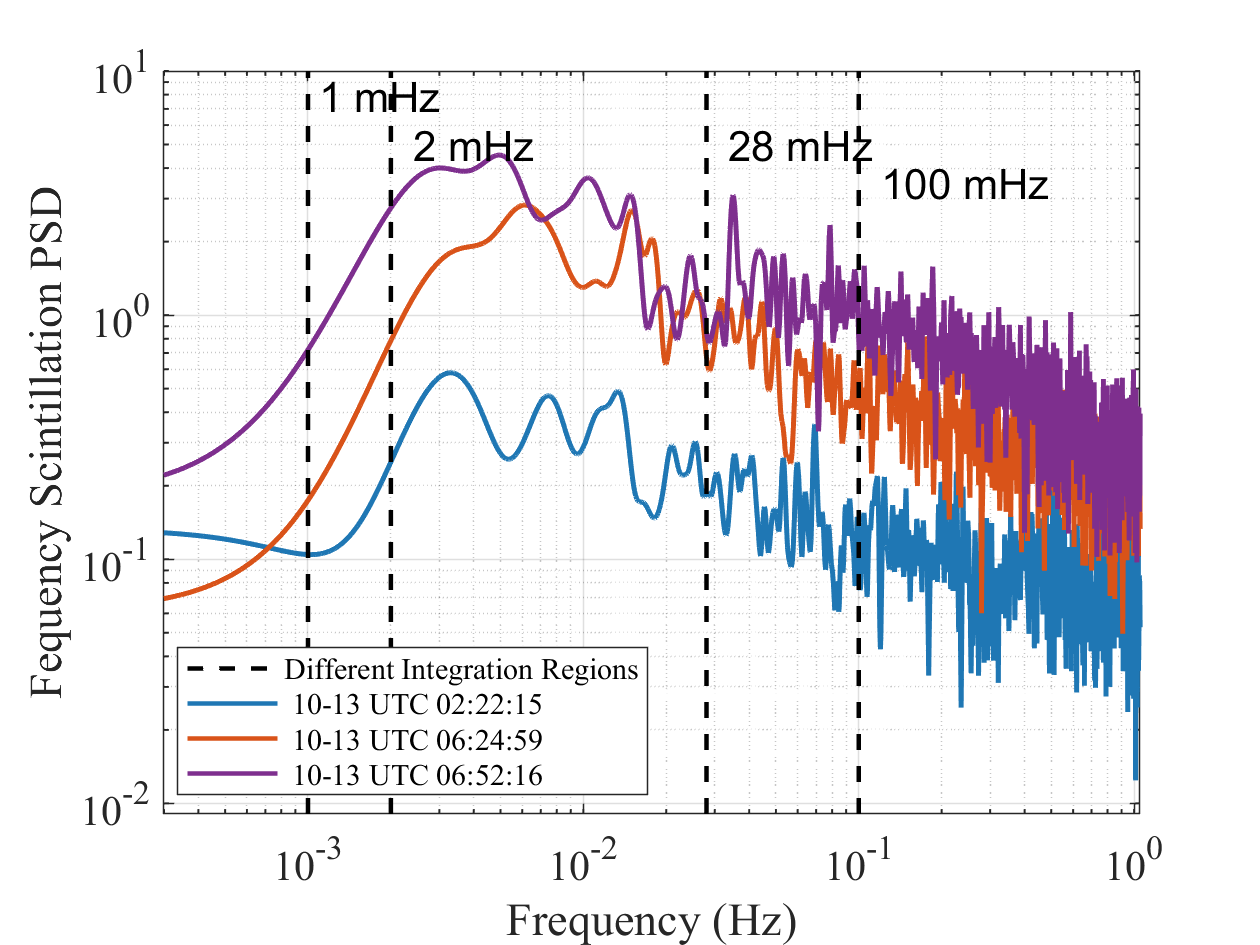}
        \caption{}
        \label{subfig:1013PSD}
    \end{subfigure}
    \hfill
    \begin{subfigure}{0.48\textwidth}
        \centering
        \includegraphics[width=\textwidth]{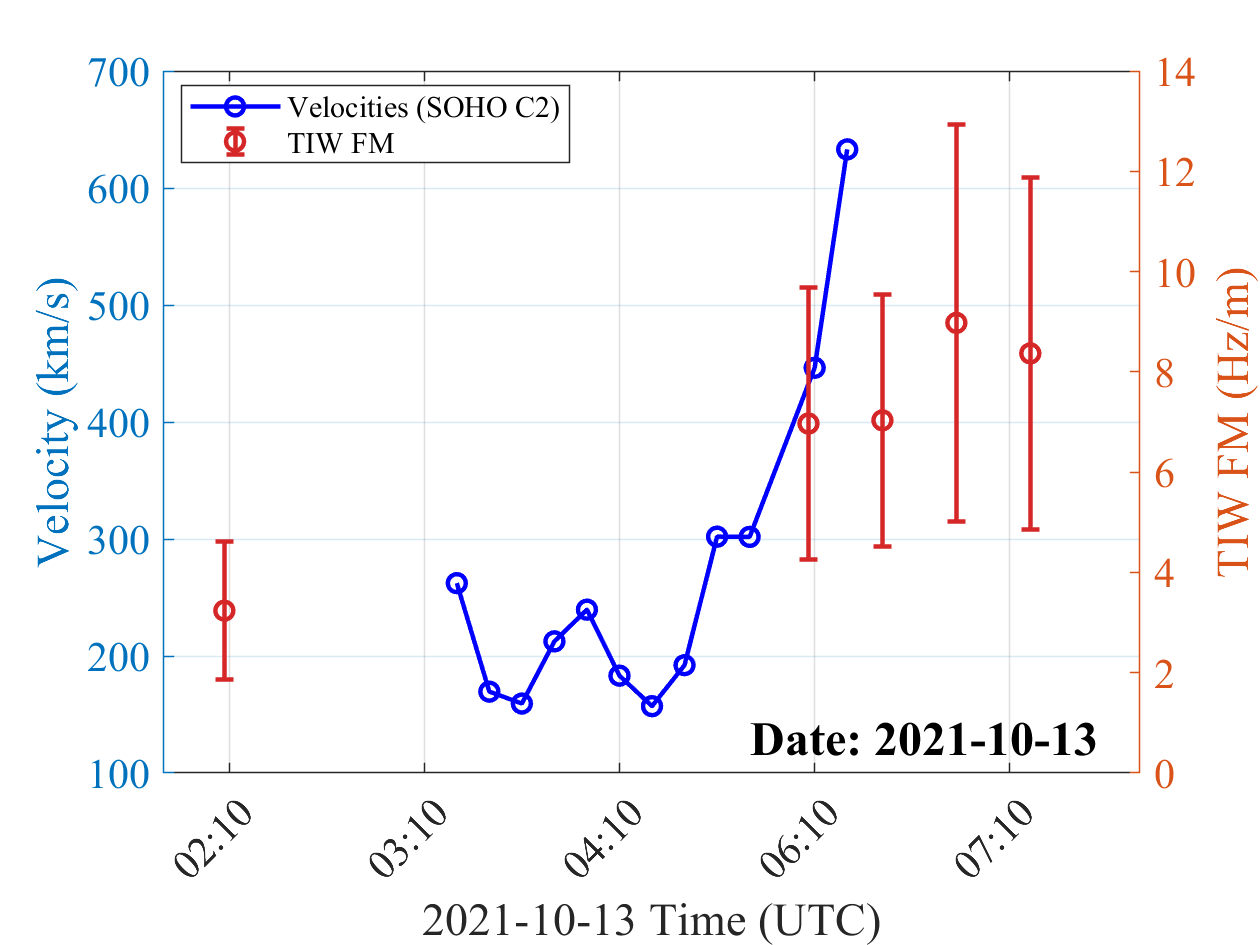}
        \caption{}
        \label{subfig:1013FM}
    \end{subfigure}
     \hfill
  \begin{subfigure}{0.8\textwidth}
        \centering
        \includegraphics[width=\textwidth]{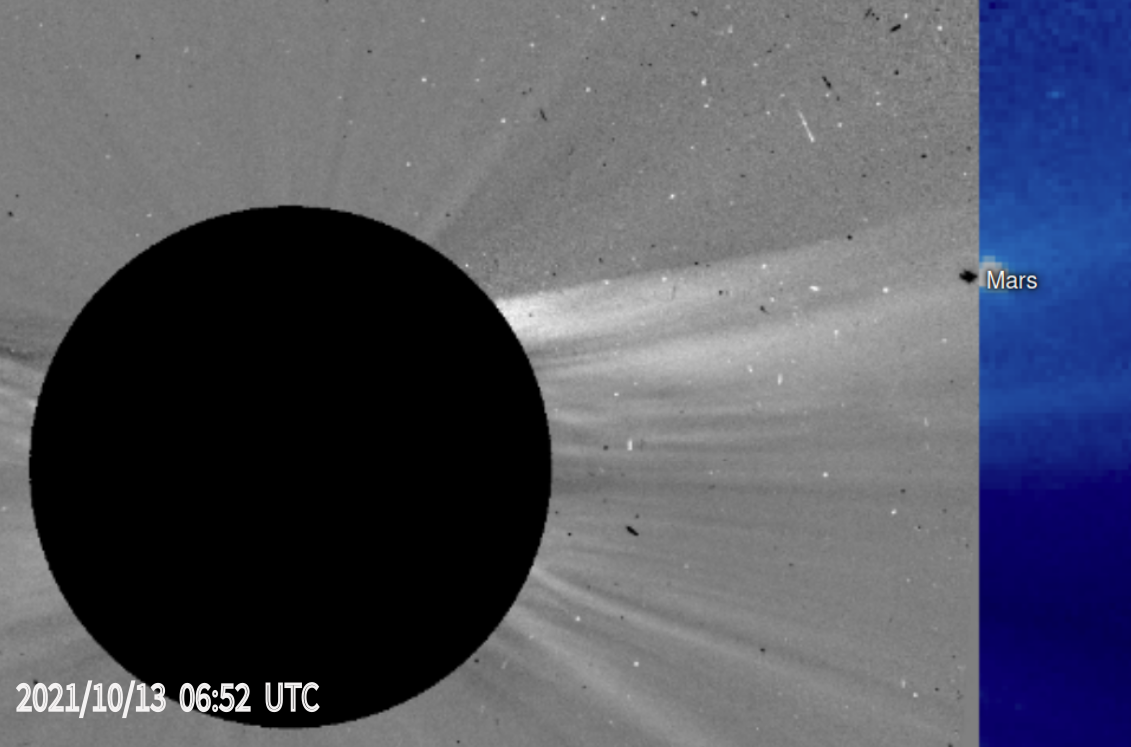}
        \caption{}
        \label{subfig:SOHO1013}
    \end{subfigure}
    \caption{ Relationship between FM fluctuations  $\sigma_{FM}$ and High-Speed Solar Wind Streams.(a):The frequency scintillation power spectrum observed on October 13.
    (b):Comparison between solar wind speed estimated from SOHO C2 and   $\sigma_{FM}$ variations.
    (c):From the inner to the outer region, the image is a composite of the SOHO C2 coronagraph and the SOHO C3 coronagraph.}
    \label{fig:1013 total}
\end{figure}

\;\;\;\;\ \
 $\sigma_{FM}$ can also serve as a means of 
detecting high-speed solar wind. 
In the outer region of the mid-corona 
(within about $6 R_{\odot}$), 
solar wind with velocities 
approaching $450 km \cdot s^{-1}$ is generally 
considered high-speed solar wind, 
while slow solar wind has velocities 
around $150 km \cdot s^{-1}$. \citep{MidCorona2023}
Therefore, the criterion can be 
simplified by considering solar wind 
exceeding $400 km \cdot s^{-1}$  as high-speed.
It is generally accepted that 
high-speed solar wind originates from coronal holes.
In the lower panel of Figure ~\ref{fig:1013 total}, 
observational data from STEREO-A EUVI 304 $\AA$ 
are included. A coronal hole can be seen 
in the upper right region of the Sun, 
extending outward and connecting to the SOHO C2 image.

The black-and-white image from SOHO C2 in Figure ~\ref{subfig:SOHO1013} 
is presented using the Base Difference method, 
which differs from the representation used earlier. 
Base Difference is one of the 
most commonly used techniques for 
background subtraction and transient enhancement. 
Its principle involves subtracting, 
pixel by pixel, a fixed reference frame 
(e.g., a quiet corona image prior to the event) f
rom the current frame. 
This suppresses slowly varying background structures, 
such as the evolution of streamers, 
and highlights transient features relative 
to the quiet Sun state.

In the difference image, the brightness 
of each pixel is determined by its value: 
values greater than 0 indicate regions 
brighter than the reference frame 
and appear as white or light gray; 
values less than 0 indicate darker regions 
and appear as black or dark gray;
values near 0 signify no significant 
change and appear as an intermediate gray.

Based on Figures ~\ref{subfig:1013PSD} and ~\ref{subfig:1013FM}, 
the solar activity at the O point corresponding 
to the first   $\sigma_{FM}$ observation of the day was relatively quiet. 
Therefore, using this observation time (2021-10-13 02:22 UTC) 
as the reference, a base difference image was generated from 
the SOHO C2 data, yielding the left panel of Figure ~\ref{subfig:SOHO1013}.

The SOHO C3 image appended to the outer side 
of the SOHO C2 image is intended to indicate 
the position of Mars. 
The white pixels on the right side of the 
C2 image indicate the presence of a solar wind
jet near 06:52 UTC, 
which lies exactly along the line of 
sight between SOHO and Mars.

Calculations indicate that 
the velocity of this jet around 06:00 UTC
had already exceeded $400 km \cdot s^{-1}$,
classifying it as high-speed solar wind. 
The temporal evolution of its velocity 
is summarized in Figure ~\ref{subfig:1013FM}.

\subsection{Coronal Streamer}
\begin{figure}[htbp]
    \centering
    \begin{subfigure}{0.51\textwidth}
        \centering
        \includegraphics[width=\textwidth]{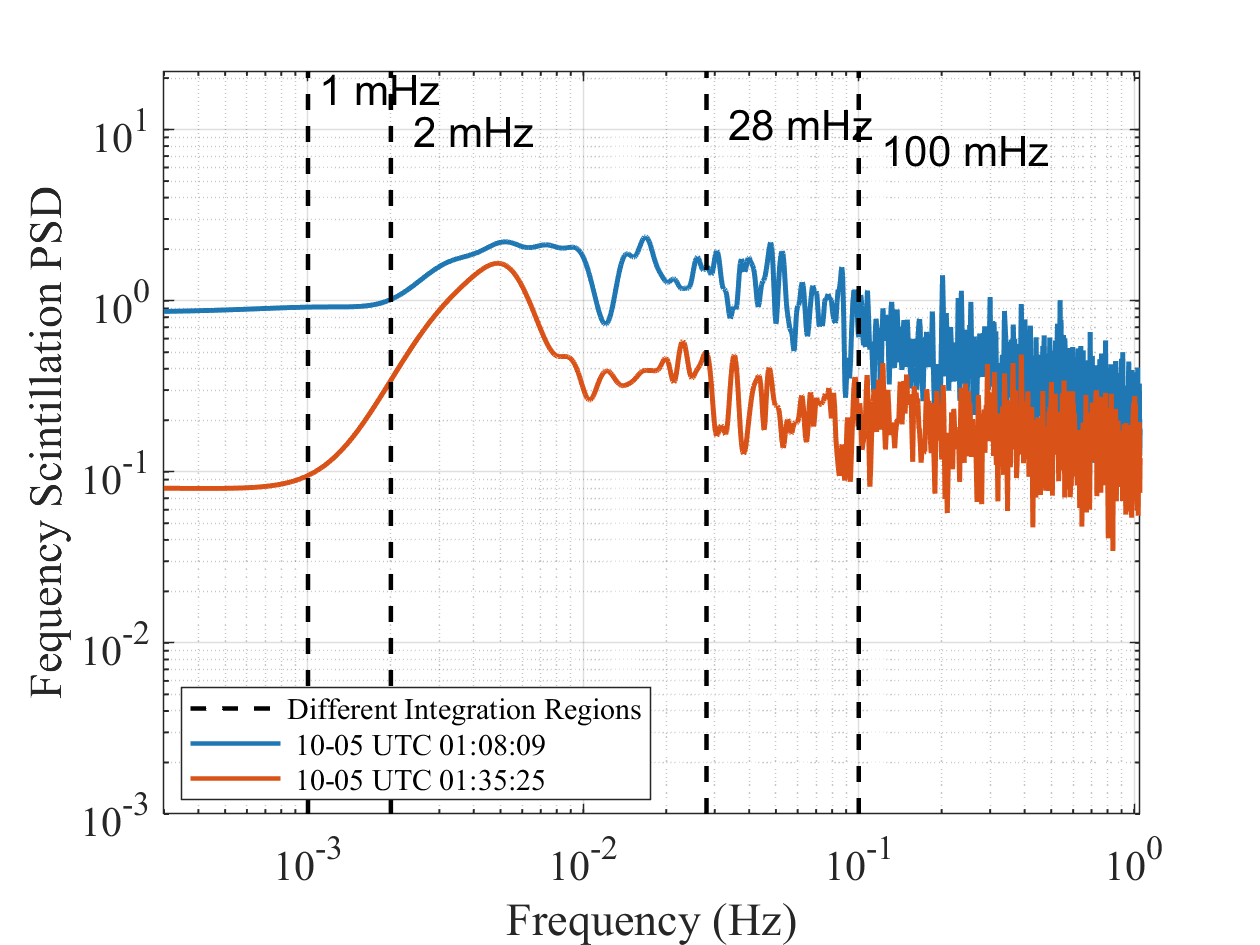}
        \caption{}
        \label{subfig:1005PSD}
    \end{subfigure}
    \hfill
    \begin{subfigure}{0.48\textwidth}
        \centering
        \includegraphics[width=\textwidth]{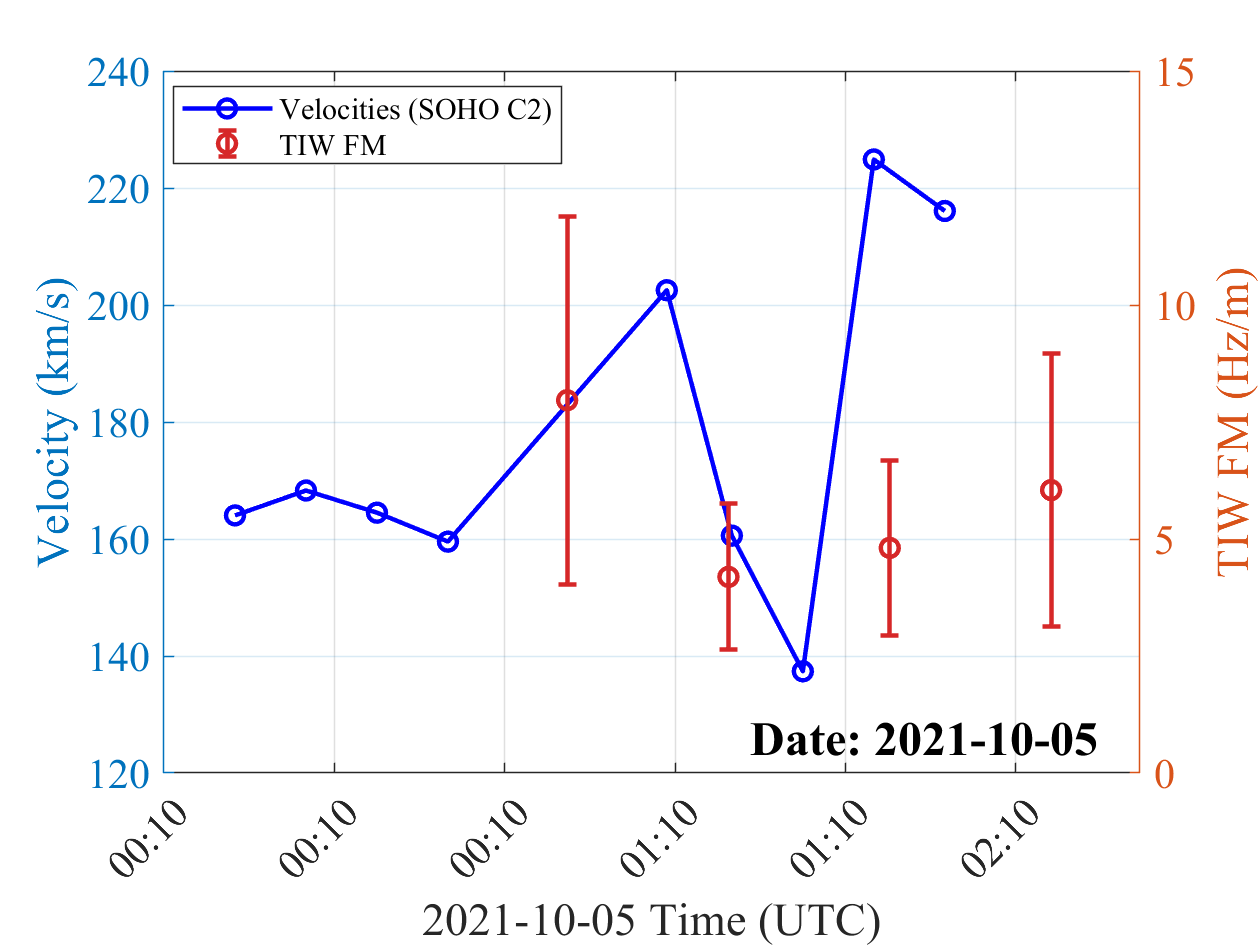}
        \caption{}
        \label{subfig:1005FM}
    \end{subfigure}
     \hfill
  \begin{subfigure}{0.8\textwidth}
        \centering
        \includegraphics[width=\textwidth]{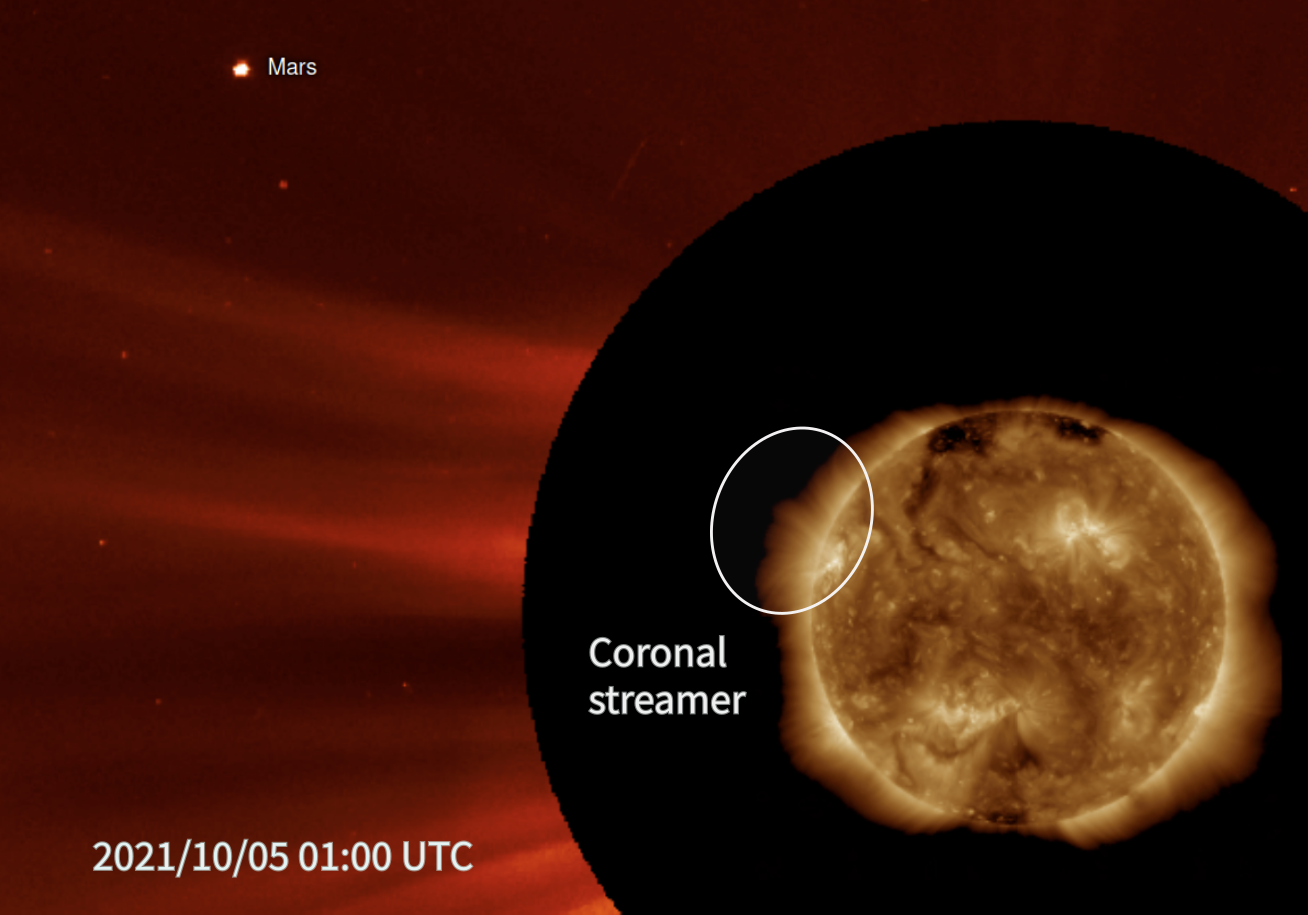}
        \caption{}
        \label{subfig:SOHO1005}
    \end{subfigure}
    \caption{ Relationship between FM fluctuations  $\sigma_{FM}$  and Coronal Streamer.
     (a):The frequency scintillation power spectrum  observed on October 05.
    (b):Comparison between solar wind speed estimated from SOHO C2 and FM variations  $\sigma_{FM}$. 
    (c):From the inner to the outer region, the image is a composite of SDO AIA 193 \AA, the SOHO C2 coronagraph.}
    \label{fig:1005 total}
\end{figure}

\;\;\;\;
The lower panel of Figure ~\ref{fig:1005 total} 
incorporates observational data from the SDO AIA 193 \AA\, 
channel. 
The 193 \AA \, channel is most sensitive 
to coronal plasma at approximately 
1.5 million K, and coronal streamers 
happen to be high-density, 
high-brightness structures at this temperature, 
thereby achieving the highest imaging 
contrast and clearest morphology. 
This waveband simultaneously highlights 
the bright structure of streamers 
under closed magnetic fields 
and the distinct brightness 
contrast with coronal holes 
under open magnetic fields, 
making it the optimal band for 
observing the position, extent, and structure of streamers.
 
A distinct coronal streamer can be clearly 
observed extending from the Sun's upper-left region, 
connecting outward to the features captured in the 
SOHO C2 image. 
Notably, the direction of this streamer's eruption 
aligns closely with the position of Mars, 
indicated as point $O$. 
Coronal streamers are typically associated 
with the generation of slow solar wind, 
a correlation that is effectively validated 
by the data presented in Figure ~\ref{subfig:1005FM}.

\subsection{Spatial Accuracy of Anomaly Determination}

The preceding sections have verified that   $\sigma_{FM}$ anomalies 
can be used to identify the presence of CMEs, 
high-speed solar wind, and coronal streamers, 
with good temporal correspondence in the data. 
The next step is to verify whether    $\sigma_{FM}$anomalies 
can accurately locate anomalous solar activity in space. 
A counterexample will be presented to illustrate this.

\begin{figure}[htbp]
    \centering
    \begin{subfigure}{0.51\textwidth}
        \centering
        \includegraphics[width=\textwidth]{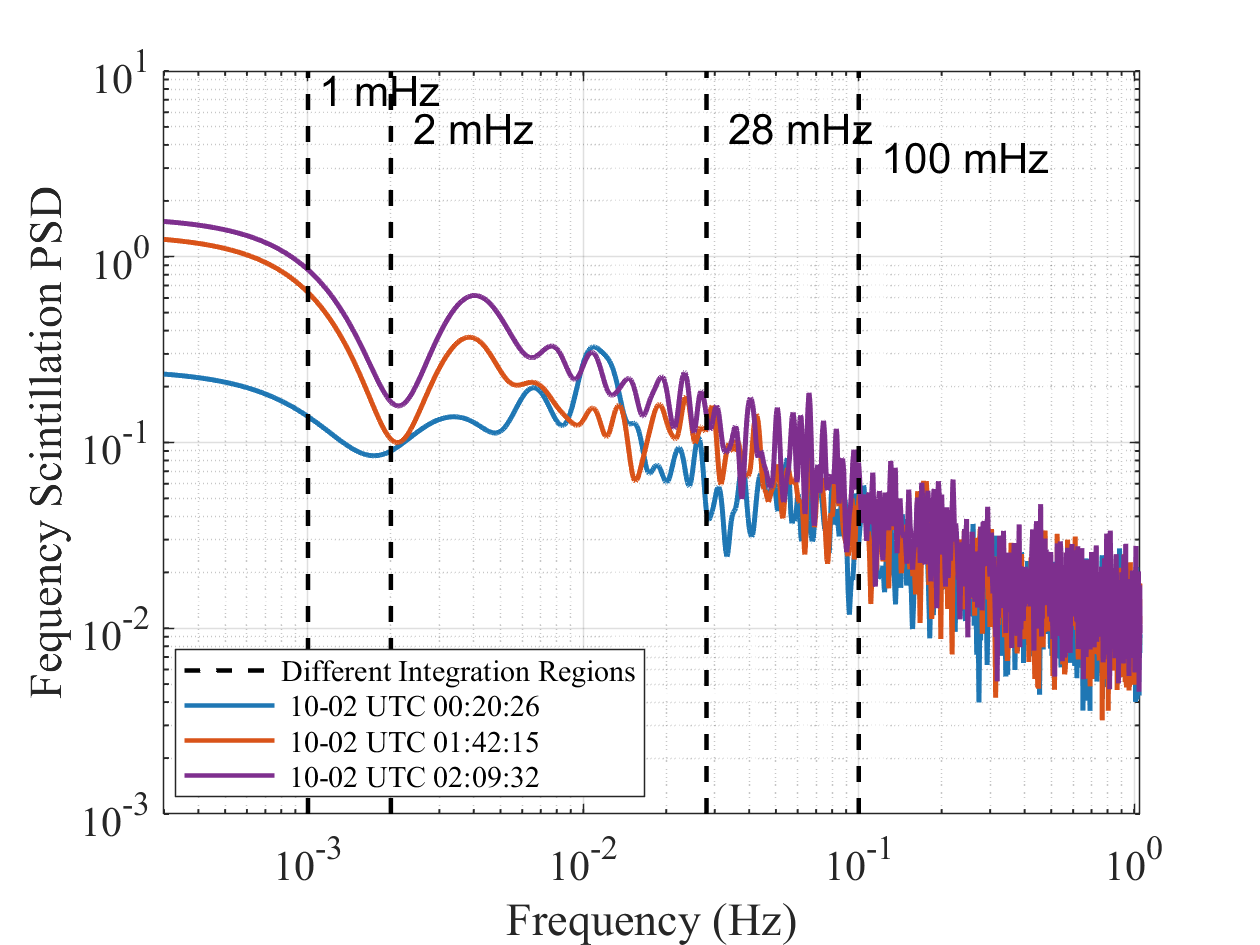}
        \caption{}
        \label{subfig:1002PSD}
    \end{subfigure}
    \hfill
    \begin{subfigure}{0.48\textwidth}
        \centering
        \includegraphics[width=\textwidth]{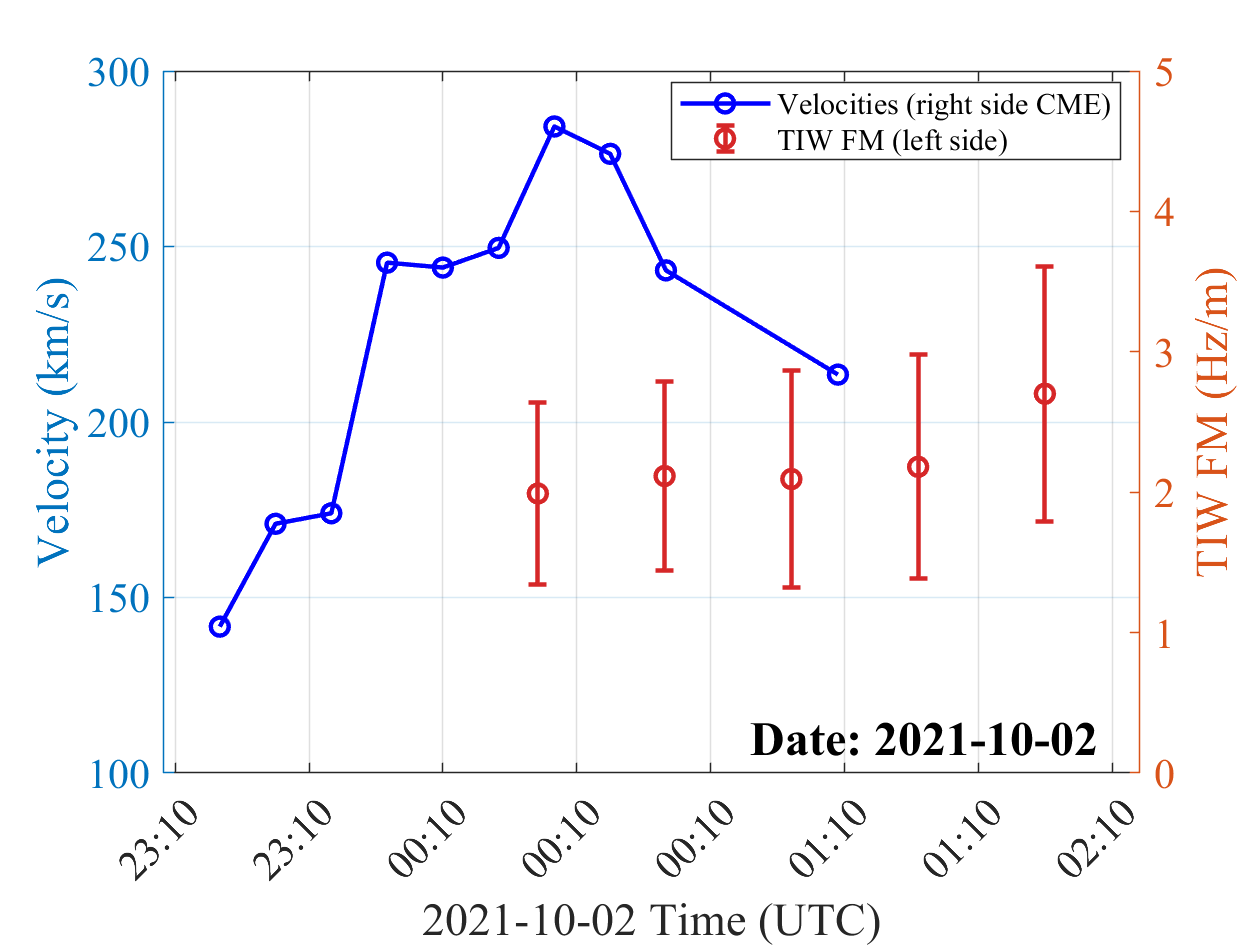}
        \caption{}
        \label{subfig:1002FM}
    \end{subfigure}
     \hfill
  \begin{subfigure}{0.8\textwidth}
        \centering
        \includegraphics[width=\textwidth]{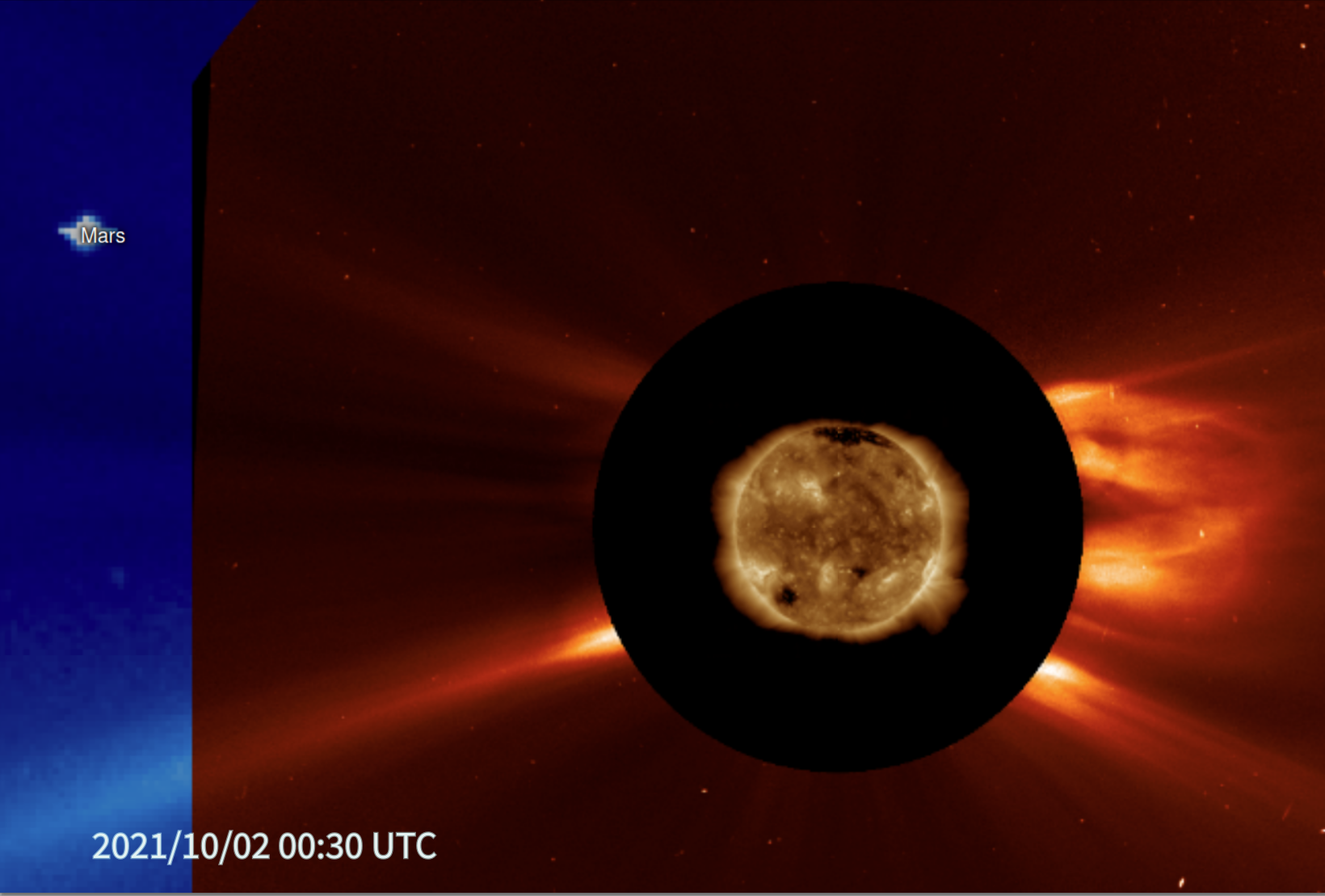}
        \caption{}
        \label{subfig:SOHO1002}
    \end{subfigure}
    \caption{Relationship between   $\sigma_{FM}$ and spatially mismatched CMEs.
    (a):The frequency scintillation power spectrum observed on October 02.
    (b):Comparison between solar wind speed estimated from SOHO C2 and FM variations $\sigma_{FM}$.
    (c):From the inner to the outer region, the image is a composite of SDO AIA 193 \AA, the SOHO /LASCO C2 coronagraph and the SOHO C3 coronagraph.
    }
    \label{fig:1002_total}
\end{figure}

First, as shown in Figure ~\ref{subfig:1002PSD}, 
the theoretical scintillation spectrum 
exhibits either a flat trend in the low-frequency band 
or a descending pattern after a slight rise \citep{efimovOuterScaleSolarwind2002}.
This once again confirms that the scintillation spectrum 
below 2 mHz may be distorted due to the 
fitting accuracy of the Doppler trend term, 
thereby validating the effectiveness of 
using 2 mHz as the lower integration limit.

As can be seen from both Figure ~\ref{fig4:sub1} 
and Figure ~\ref{fig:solar_radii_FM},
the $\sigma_{FM}$ on 2021 October 2 does not exhibit 
any anomaly. 
First, the arrangement of the 
frequency scintillation spectrum follows the pattern of 
gradually increasing over time during the ingress phase. 
Moreover, the $\sigma_{FM}$ shows no abnormal 
increase relative to Equations (~\ref{eqs:sigma_FM_eq1}) 
and (~\ref{eqs:sigma_FM_eq2}), 
a point that is also clearly evident in Figure ~\ref{fig:solar_radii_FM}.

However, images from SOHO LASCO C2 show that on that day, 
a strong CME eruption occurred on the right side of the Sun, 
and the eruption process lasted throughout 
the period of the frequency scintillation observation experiment. 
Combining this with the previous results, 
it can be inferred that since the CME 
did not pass through point $O$, 
even a strong eruption in space would 
not affect the downlink signal, 
and thus no anomaly would be observed in the $\sigma_{FM}$. 
This observational result demonstrates 
that anomalies in $\sigma_{FM}$ are strongly 
correlated with the spatial distribution of solar activity.

\subsection{Quantitative analysis of time delays}
\label{sec:corr_velo}

\;\;\;\;\ \

In this section, we quantitatively 
analyze the time offsets between 
the frequency fluctuation variance $\sigma_{FM}$ anomalies 
and the SOHO-inferred 
solar wind speeds across multiple events, 
with a detailed focus on the two events on October 13 and 15.
We adopt a consistent quantitative analysis framework 
to verify the reliability of the temporal correlation 
between the $\sigma_{FM}$ anomaly peaks and the 
SOHO-inferred solar wind speeds.

For the two events on October 13 and 15, 
we conducted the research following a 
consistent analytical approach. 
As shown in Figure ~\ref{subfig:1015FM}
(Comparison of $\sigma_{FM}$ and Solar Wind Speed Variations), 
there is a significant time lag 
between the $\sigma_{FM}$ anomaly peaks 
and the solar wind speed anomalies observed by SOHO: 
in the October 15 event, 
the occurrence time of the $\sigma_{FM}$ 
peak was later than that of the solar wind speed anomaly; 
as displayed in Figure ~\ref{subfig:1013FM}, 
the October 13 event also exhibited this characteristic, 
with a distinct 
lag of the $\sigma_{FM}$ peak relative to the solar wind speed anomaly.

Figure ~\ref{subfig:1015FM} demonstrates that 
$\sigma_{FM}$ 
achieves its maximum value at approximately 
06:14 UTC, while SOHO observations identify the
 CME peak speed between 04:48 and 05:36 UTC.

This yields an observed time delay of approximately 
38 minutes to 1 hour and 26 minutes 
for the October 15 event.
For the high-speed solar wind event on October 13, 
as shown in Figure ~\ref{subfig:1013FM}, the $\sigma_{FM}$ anomaly peak appears at 06:52 UTC, 
and the period of significant enhancement in the SOHO solar wind speed 
spans from 06:00 to 06:12 UTC. 
Consequently, the observed time delay for this event 
is estimated to be around 40 minutes to 52 minutes.

As illustrated in Figure ~\ref{fig:IPS_theory}
(Schematic Diagram of Signal Propagation Path), 
the  point $O$ — the location 
where electromagnetic waves are most strongly disturbed — 
is roughly located between Mars 
and Earth and represents the closest observational position 
to the Sun. Consequently, the propagation time delay of 
the perturbed signal can be approximated as half of
the total communication time delay between Earth and Mars. 
For the event on October 15, the observation point $O$ 
is approximately 9 solar radii away from the heliocenter. 
In contrast, the solar wind features observed 
by SOHO on that day originated from a region 
with a heliocentric distance ranging from 5.5 to 6.5 solar radii. 
Thus, the radial distance between the solar wind source 
region on the SOHO images and the observation point $O$ is 
calculated as follows:

\begin{equation}
\Delta r_{15} = 9R_\odot - (5.5\sim6.5)R_\odot = 2.5\sim3.5R_\odot \approx 1.74\times10^6\sim2.43\times10^6\ \text{km},
\end{equation}

Here,  $9R_\odot \approx 695700$ km denotes the solar radius.
Assuming that the disturbance propagates uniformly along the radial
direction, the propagation time of the solar wind disturbance 
can be estimated by the following formula:

\begin{equation}
t = \frac{\Delta r}{v},
\end{equation}

In the formula, $v$ represents the propagation speed of the CME. 
For the October 15 event, 
we adopted the observed CME speed ($550\sim 600 km\cdot s^{-1}$)
 and combined it with the corrected propagation distance 
 (corresponding to the heliocentric distance of $5.5\sim6.5R_\odot $), 
 and the calculated expected 
propagation time for the October 15 event is:

\begin{equation}
t_{15} \approx \frac{1.74\times10^6\ \text{km}}{600\ \text{km/s}} \sim \frac{2.43\times10^6\ \text{km}}{550\ \text{km/s}} \approx 0.79\sim4.42\ \text{hours}.
\end{equation}

For the event on October 13, 
point O is located at approximately 
6.5 solar radii from the heliocenter, 
while the solar wind features observed by 
SOHO on that day originated from a 
heliocentric distance of 3.5–4.5 solar radii. 
The radial distance from the solar wind source region in the 
SOHO images to observation point $O$ is therefore:
\begin{equation}
\Delta r_{13} = 6.5R_\odot - (4\sim4.5)R_\odot = 2\sim2.5R_\odot \approx 1.39\times10^6\sim17.4\times10^6\ \text{km},
\end{equation}

Using the observed solar wind speed of $500 km\cdot s^{-1}$ 
for the October 13 event, 
the expected propagation time of the solar wind 
disturbance is calculated as:
\begin{equation}
t_{13} = \frac{\Delta r_{13}}{v_{13}} \approx \frac{1.31\times10^6\ \text{km}}{500\ \text{km/s}} \sim \frac{1.66\times10^6\ \text{km}}{500\ \text{km/s}} \approx 0.73\sim0.92\ \text{hours}.
\end{equation}

Accounting for an additional signal propagation 
time of approximately 20 minutes from Mars to Earth, 
we may simply take half of this 20‑minute interval, 
i.e., 10 minutes ($\approx0.17$ hours), as the propagation 
time passing point $O$. 
After incorporating this signal propagation delay, 
the total time delay for the October 13 event 
is approximately 0.9–1.09 hours, 
which is in good agreement with the observed delay 
of around 0.87 hours. For the October 15 event, 
the total time delay is about 0.96–4.59 hours,
consistent with the observed delay range.
In addition, the solar wind propagation velocity 
calculated from coronagraph optical images may also 
be subject to certain deviations due to the effects of 
image resolution and noise.

This consistency further corroborates the physical 
connection between $\sigma_{FM}$ anomalies 
and solar wind disturbances, demonstrating 
that $\sigma_{FM}$  anomalies are a direct manifestation 
of solar wind disturbances propagating to the observation point.
It also fully validates the rationality of the quantitative 
analysis method, providing reliable analytical 
approaches and data support for relevant follow‑up studies.

\section{CONCLUSIONS}\label{sect:conclusion}
\;\;\;\;\ \
The observation experiment of solar wind parameter 
retrieval was carried out by 
using the downlink data of Tianwen 1 probe received 
by WRT70
during the superior conjunction of Mars in 2021. 
During the experiment, the reliability of 
the multi-level iteration correction algorithm 
of Doppler long-term trend removal was verified 
under the interference of coronal activity. 
This method enables a more accurate estimation 
of Doppler frequency scintillation under strong signal interference, 
making it suitable for communication environments 
with severe interference during solar occultation. 
It can maximally extract the low-frequency components 
of the frequency scintillation spectrum from the downlink signal, 
thereby reflecting variations in the large-scale structure 
as the signal propagates through regions of solar activity. 
It has been validated using multi-day data that, 
under the current data processing pipeline, 
the effective accuracy at the lowest frequency band can reach 2 mHz.

The inversion of Doppler scintillation spectra 
within 10 solar radii and the estimation and 
analysis of the normalized RMS 
frequency fluctuation measure $\sigma_{FM}$ are carried out. 
In this paper, the statistics of the amount of
Doppler scintillation during the superior conjunction period
is obtained, and it is found that the results 
are in good agreement with the observed empirical formula, 
and the statistical results reveal the turbulence characteristics 
of the solar wind at the interplanetary scale. 
According to the experimental results, 
the turbulence spectra of the solar wind within 10 solar radii 
generally meet the rule that the normalized RMS frequency fluctuation 
measure $\sigma_{FM}$ increases with the decrease of the 
SO distance.

It is found that the Doppler scintillation spectra on 
2021 October 5, 13, and 15 
 have the characteristics 
of rising in the low frequency band of the Frequency Scintillation PSD 
and $\sigma_{FM}$ with the increase of SO distance.
By comparing the observational data 
from the SOHO coronagraph in the optical band, 
it was found that the anomalies over these three days 
were all caused by solar activity, 
albeit for slightly different reasons. 
The anomalous $\sigma_{FM}$  variation patterns 
during these three days are attributed to 
coronal streamers, high-speed solar wind, 
and coronal mass ejections (CMEs), respectively.

By analyzing the solar wind images observed by SOHO, 
the variations in solar wind velocity 
can be obtained and compared with the variations 
in $\sigma_{FM}$. The results indicate that 
fluctuations in $\sigma_{FM}$  are strongly 
correlated with fluctuations in solar wind velocity; 
however, there is a certain time delay in FM variations  $\sigma_{FM}$
relative to the solar wind velocity changes 
derived from coronagraph optical data. 

This time delay may be associated with two factors. 
First, the solar wind velocity derived from coronagraph 
observations requires a certain propagation time to reach point $O$ 
due to the distance effect. 
Second, the downlink signal 
takes approximately 20 minutes to travel from Mars back to Earth, 
and an additional time delay is introduced as 
the signal passes point $O$  before being received on the ground. 
Together, these two factors give rise to a certain time 
lag in the temporal correlation between solar wind velocity 
and $\sigma_{FM}$ variations.
Through relevant calculations, 
we verified the rationality of the quantitative analysis 
method and confirmed that the observed time lag between 
the $\sigma_{FM}$ 
fluctuation peaks and solar wind velocity anomalies 
on October 13 and 15 is at least about 50 minutes. 
Simple theoretical analysis shows that the theoretical
time delay agrees well with the experimental measurements,
and the total time delay accounting for the signal propagation 
effect is highly consistent with the actual observations. 
This quantitative verification provides reliable data support 
for further in-depth research on solar wind 
disturbances and their electromagnetic effects.

The results presented in this paper not 
only validate the temporal correlation between $\sigma_{FM}$ 
variations and coronal activity but also demonstrate, 
through a counterexample, the spatial correlation 
between $\sigma_{FM}$ variations and coronal activity.
Before discussing the counterexample, 
it should be emphasized that the anomalies observed 
on October 5, 13, and 15 
have all been confirmed 
by SOHO imagery to be associated with 
coronal streamers, high-speed solar wind, 
and CMEs that crossed point $O$. 
The situation on October 2, however, 
provides a perfect counterexample. 
On that day, the frequency scintillation observation 
data showed no anomalies and followed the pattern of $\sigma_{FM}$ steadily 
increasing as SO decreased. 
However, during the observation period 
covered by the experiment on October 2, 
a strong CME eruption occurred on the right side of the Sun 
from the perspective of SOHO, 
while the downlink signal path was located on the left side of the Sun. 
This result demonstrates that $\sigma_{FM}$ anomalies 
are also related to the spatial distribution of the signal path relative 
to coronal activity.

\section{Future direction}
\;\;\;\;\ \
With the development of China's deep space exploration mission and manned space program, the study of solar wind and the construction of prediction models have become an important issue. It can be expected that more and more observations will be made near the sun, and the next stage of research will focus on the following aspects.
\begin{enumerate}
    \item In this study, we have so far conducted 
    a qualitative analysis of coronal activity using $\sigma_{FM}$
    and preliminarily determined that fluctuations in $\sigma_{FM}$ may be associated with 
    high-speed solar wind, coronal streamers, and CMEs
    The next step aims to further distinguish the distinct 
    effects of these three types of coronal activity 
    on electromagnetic signals, 
    thereby enabling a more precise determination 
    of the type of coronal activity based on FM variations  $\sigma_{FM}$
    in future studies;
    \item Using the observation data of different observation bands, 
    combined  with new equipment such as 
    Parker Solar Probe(PSP), Chinese H$\alpha$ Solar Explorer(CHASE) 
    and Mingantu IPS (Interplanetary Scintillation) telescopes, the correction and modeling 
    of the intensity scintillation inversion method 
    in the near-solar region are completed;
    \item  By utilizing $\sigma_{FM}$ to detect the 
    spatial correlation of coronal activity and 
    combining observational results from multiple stations, 
    it is possible to perform three-dimensional localization 
    and reconstruction of coronal activity, 
    thereby more accurately delineating the extent of 
    solar active regions;
    \item The communication channel between the 
    Earth and the Mars is modeled by using the 
    downlink data during the superior conjunction of Mars. 
    Further study on the influence of solar wind on 
    communication link and communication compensation technology.
\end{enumerate}

\begin{acknowledgements}
We thank the anonymous reviewers for their valuable comments,
which have greatly helped in improving this manuscript.    
This work was funded by the Astronomical Joint Fund of the
National Natural Science Foundation of China and Chinese
Academy of Sciences (Grant Nos. 12373103).
\end{acknowledgements}

\appendix                  

\label{lastpage}

\end{CJK}

\end{document}